\documentclass[aip,pof,reprint, twocolumn]{revtex4-1}
\usepackage{graphicx}
\usepackage{dcolumn}
\usepackage{float}
\usepackage{xfrac}
\usepackage{amsmath}
\usepackage{hyperref}
\usepackage{romannum}
\usepackage{amsfonts}
\newcommand{\bigepsilon}{\makebox{\large\ensuremath{\epsilon}}}

\usepackage{upgreek}

\usepackage{relsize}
\usepackage{bm}
\usepackage{setspace}
\usepackage{multirow}
\begin{document}
\pagenumbering{arabic}

\title{Dynamics and Statistics of Reorientations of Large-Scale Circulation in Turbulent Rotating Rayleigh-B\'{e}nard Convection}
\author{Vishnu Venugopal T}

\author{Arnab Kumar De}
\affiliation{Department of Mechanical Engineering, Indian Institute of Technology Guwahati, Assam, India-781039}

\author{Pankaj Kumar Mishra}
\affiliation{Department of Physics, Indian Institute of Technology Guwahati, Assam, India-781039}

\begin{abstract}
We present a direct numerical simulation to investigate the dynamics and statistics of reorientations of large-scale circulation (LSC) in turbulent rotating Rayleigh-B\'{e}nard convection (RRBC) for air  ($Pr=0.7$) contained in a cylindrical cell with unit aspect ratio. A wide range of rotation rates ($0\leq Ro^{-1}\leq 30$) is considered for two different Rayleigh numbers $Ra=2\times10^6$ and $2\times10^7$. Using the Fourier mode analysis of time series data obtained from the different probes placed in the azimuthal direction of the container at the mid-plane, the orientation and associated dynamics of LSC are characterized. The amplitude of the first Fourier mode quantifies the strength of LSC and its phase $\Phi_1$ gives the information of the azimuthal orientation of LSC. Based on the energy contained in the Fourier modes different flow regimes are identified as the rotation rate is varied for a given Rayleigh number. LSC structure is observed in the low rotation regime ( $Ro^{-1} \lesssim 1$). A strong correlation between the orientation of LSC structure and the heat transfer and boundary layer dynamics is observed. In the LSC regime, the dissipation rates follow the log-normal behaviour, while at higher rotation rates, a clear departure from log-normality is noticed. Different types of reorientations, namely, rotation-led, cessation-led, partial and complete reversal are identified.
The distribution of change in orientation of LSC follows a power law behaviour as $P(|\Delta \Phi_1|) \propto |\Delta \Phi_1|^{-m}$, with the exponent $m\approx 3.7$.  In addition, the statistics of time interval between successive reorientations follow a Poisson distribution. These observations are in good agreement with earlier experimental results.

\end{abstract}


\maketitle 

\vspace{3 mm}
\section{Introduction}
\par
Thermal convection has been of major interest to the scientific community owing to its ubiquitous presence in many natural flows and engineering applications. Rayleigh-B\'{e}nard convection (RBC), in which a thin fluid layer is confined between a hot isothermal plate at the bottom and cold isothermal plate at the top, is an ideal prototype model to study convective turbulence. It is one of the few systems in which the flow approaches to the turbulent state upon increase in the parameter quite systematically, i.e., static region is followed by the time dependent region and then chaos which is followed by the spatio-temporal chaos and then turbulence \cite{Krishnamurthi_Howard_1981}.   Flow characteristics of RBC depend on  
the Rayleigh number $Ra=g\beta \Delta T H^{3}/\nu \alpha$, the Prandtl number $Pr = \nu/\alpha$ and the aspect ratio $~\Gamma=D/H$, the ratio of the horizontal to the vertical dimension of the container.
Where $\Delta T$ is the temperature difference between the bottom and top plates, $g$ the acceleration due to gravity, and $\alpha$, $\beta$ and $\nu$ are respectively the thermal diffusivity, thermal expansion coefficient and kinematic viscosity of the fluid.
The other parameters which are important are the Nusselt number $Nu$, defined as the ratio of the total heat flux to conductive heat flux and the Rossby number $Ro=V/2\Omega H$ (in case of rotating RBC), where $V$ is the free-fall velocity and $\Omega$ is the rotation rate.
\smallskip
\par
In turbulent state of RBC above a threshold Rayleigh number statistical coherence of the flow is restored with the appearance of large-scale circulation also known as ``mean-wind'' with turbulent background which is a self-organized structure having length scale of order of size of the container and time-scale of the order of eddy turn-over time.\cite{Krishnamurthi_Howard_1981,Casting_etal_1989,Sano_Wu_1989,Ciliberto_etal_1996,Qiu_Tong_2001a,Sun_etal_2005a,Sun_etal_2005b,Tsuji_etal_2005,Brown_Ahlers_PRL_2007}
On the course of time, the plane containing LSC exhibits change in its orientation also known as reorientations of LSC, which can occur in two ways in cylindrical domain: (a) by an azimuthal rotation of the plane containing LSC  without considerable change in the circulation strength \cite{Cioni_etal_1997,Brown_Nikolaenko_Ahlers_PRL_2005}, called as rotation-led reorientation. (b) by momentary vanishing of the circulation strength accompanied by arbitrary change in the orientation of flow, termed as cessation-led reorientation \cite{Hansen_etal_1992,Brown_Nikolaenko_Ahlers_PRL_2005}. 
Reorientations in which the large-scale flow completely changes its direction of circulation are termed as flow reversals \cite{Cioni_etal_1997}. Azimuthal changes in the orientation of LSC has been previously observed by many researchers \cite{Keller_1966, Welander_1967, Creveling_etal_1975, Gorman_etal_1984, Brown_Ahlers_JFM, Sreenivasan_etal_2002, Xi_Xia_PRE_2008}. Sudden changes in the orientation of LSC are not only interesting phenomenon, but are extremely relevant in many of the geophysical and astrophysical flows. For example, field reversals are common in the natural convection of Earth's atmosphere \cite{Van_Doorn}, magnetic field reversals in geodynamo \cite{Mishra_etal_EPL_2013}, etc.
Understanding the genesis of LSC and the nature of the instabilities that lead to the reorientations of LSC is one of the classical problems in RBC \cite{Ahlers_Grossmann_Lohse}. 
\par
Experiments have played a key role in understanding the characteristics and dynamics of LSC in turbulent RBC.
Cioni \emph{et al} \cite{Cioni_etal_1997} identified LSC from dipolar temperature distribution obtained from probes placed along the azimuth of the container. Reversals in the flow were also obtained as the fluctuations in temperature switched sign randomly. They computed the Fourier transform of the temperature field along the azimuthal direction and observed that the amplitude of the first Fourier mode fluctuates about a finite value. However, the phase changes by $\pi$-radians time to time indicating the flow reversals. Niemela \emph{et al.} \cite{Niemela_etal_2001} and Sreenivasan \emph{et al.} \cite{Sreenivasan_etal_2002} also performed experiments in turbulent RBC with helium and reported reversals of LSC. Similar experiments were carried out by Brown \emph{et al.} \cite{Brown_Nikolaenko_Ahlers_PRL_2005} and Brown and Ahlers \cite{Brown_Ahlers_JFM} with water as the working fluid. The probes were placed at three different horizontal planes ($z=1/4H$, $1/2H$ and $3/4H$) to get better idea on the dynamics of LSC. They identified reorientations of LSC using the phase of the first Fourier mode $\Phi_1$, and quantified them as rotation-led and cessation-led based on the amplitude of the modes. Brown and Ahlers \cite{Brown_Ahlers_JFM} observed that rotation-led reorientations were more frequent than cessation-led reorientations. 
The rotation-led reorientations exhibited power law distribution, while cessation-led reorientations followed uniform distribution.
Further, both rotations and cessations showed Poisson distribution in time. 
They observed that the probability density function (PDF) of time interval between successive reorientations follows an exponential function. Similar observations were made by Xi and Xia \cite{Xi_Xia_PRE_2008}. However, Sreenivasan \emph{et al.} \cite{Sreenivasan_etal_2002} showed that although the PDF of time interval between successive reorientations shows exponential nature, for lower time intervals it could be well fitted with power law distribution.
Xi and Xia \cite{Xi_Xia_PRE_2008} experimentally studied the aspect ratio dependency of reorientations in turbulent RBC. They observed that in contrast to the power law distribution in unit aspect ratio container, reorientations in $\Gamma=0.5$ geometry follows an exponential distribution.  
In another study \cite{Xi_Xia_PRE_2007} they found that the reorientations occur with an order-of-magnitude more frequently in $\Gamma=0.5$ than these happen in $\Gamma=1$. Funfschilling and Ahlers \cite{Funfschilling_Ahlers_2004} observed that in addition to the azimuthal meandering of direction of circulation, the upper and lower halves of LSC are also found to undergo twisting oscillations.
\par
The numerical studies on the reorientations or reversals of LSC in RBC are rare compared to the experimental ones. Verzicco and Camussi \cite{Verzicco_Camussi_1999} performed a detailed numerical study on the effect of Prandtl number on the dynamics of convective turbulent flow inside a cylindrical cell. They noticed that there is significant drop in Nusselt number when the strength of LSC decreases for lower $Pr$ compared to higher $Pr$. 
Later Stringano and Verzicco \cite{Stringano_Verzicco_JFM_2006} studied the mean flow structure in RBC in a slender cylindrical cell and observed a single convection roll breaking into two counter-rotating rolls stacked vertically one above the other. Benzi and Verzicco \cite{Benzi_Verzicco_2008} numerically investigated the statistical properties of the large scale flow in RBC at $Pr=0.7$ and $Ra=6\times10^5$, by using an external random perturbation on the temperature field.
 A detailed numerical study on the dynamics of reorientations of LSC in RBC was carried out by Mishra \emph{et al.} \cite{MishraDe}, where they reproduced the experimentally observed rotation-led reorientations, cessation-led reorientations and double-cessations for a Rayleigh number range $6\times10^5\leq Ra\leq 3\times10^7$ which is considerably lower than the experimental counterparts.
\par
Over the years many works on LSC dynamics in 2D RBC have reported that corner-rolls play an important role in flow reversals.
Here the LSC is confined to a single plane and the complicated three-dimensional dynamics are absent \cite{Breuer_Hansen_2009,Paul_2010_Pramana,Hansen_etal_1990,Hansen_etal_1992}.
Sugiyama \emph{et al.} \cite{Sugiyama_Q2D_2010} combined both numerical and experimental analysis to evaluate the LSC dynamics in 2D/quasi-2D systems for a range of $Ra$ and $Pr$. The kinetic energy and size of the corner-rolls grow with time as a result of plume detachments from
the boundary layers, and finally they take over the main large-scale diagonal flow, thus resulting in reversal.
Similar relevance of corner-rolls in flow reversals has been reported in many numerical and experimental studies in 2D/quasi-2D systems \cite{Chandra_Verma_2011, chandra_Verma_2013,Vasilev_etal_2011,Yanagisawa_etal_2011,Wagner_Shishkina_2013,Ni_Huang_Xia_2015}.
However in 3-dimensional flow these phenomena are much complex. It is established that corner-rolls do exist in  3-D systems also \cite{Sun_etal_2005a,Foroozani_etal_2017}, but unlike the  2D/quasi-2D counterparts, these rolls are not strictly confined in the LSC plane (for 3-D systems). Resultantly, when fed with energy, the corner rolls need not essentially grow in diameter at the expense of shrinking the main LSC roll; rather, they can move or grow outside the LSC plane. De and Mishra \cite{De_Mishra_2018_IHTC} investigated the dynamics of the LSC in the turbulent RBC ($2\times10^6 \leq Ra \leq 2\times10^7$) inside a cubic box and reported that the plane containing LSC is aligned along one of the diagonal direction of the box. Further, they observed that over the course of time this plane containing LSC undergoes intermittent and chaotic switching between the two diagonals. Recent simulations by De \emph{et al.}\cite{De_EJMB_2018} have shown that in case of periodic boundary conditions, the thermal plumes drift in the lateral direction that leads to the flow reversals.
\par
Although, there are many existing literatures on the reorientation dynamics of LSC in turbulent RBC, the effect of rotation on these dynamics is less explored.
Kunnen \emph{et al.} \cite{kunnen2} investigated RRBC both numerically and
experimentally at $Ra=2\times10^9$ and $Pr=6.4$. At low rotation rates a domain-filling LSC was observed ($Ro^{-1}\lesssim 0.83$), while it breaks down at higher rotation rates. At higher rotation rates the Ekman-vortex structures diminishes the strength of the LSC, and as a result for $Ro^{-1}$ significantly above unity, the LSC structure breaks down \cite{Zhong_Ahlers_JFM_2010,Hart_etal_pof_2002}. They observed a retrograde precession of the LSC structure which depends on the rotation rate. However, prograde precession of the LSC at modest rotation rates ($Ro^{-1}\lesssim 1$) was observed by Weiss and Ahlers \cite{Weiss_Ahlers_JFM_2011_LSF} in their experimental investigations on RRBC in cylinder with aspect ratio $\Gamma=0.5$ for Rayleigh number range $2.3\times10^9\lesssim Ra\lesssim 7.2\times10^{10}$.
Further, they observed that for taller aspect ratio containers LSC can exist as single-roll or double-roll systems and modest rotation has an effect of stabilizing the single-roll and destabilizing the double roll. This was later numerically observed by Stevens \emph{et al.} \cite{stevens7} for the same parameter range.
\par
The questions which revolve around the research of the dynamics of LSC are: how the coherent structure appears amidst the turbulent background; how the heat transfer and boundary layer dynamics get affected due to its presence; what are the different characteristics of the reorientations and the corresponding distributions of occurrence of these events, etc. As discussed above there are many experimental works which have taken lead in these directions, but still comprehensive numerical and theoretical studies which may give some quantitative picture to them are lacking.  In this paper, using the direct numerical simulations of RRBC we have been able to distinguish the different flow regimes (LSC dominated or not) by calculating relative energy contained in the Fourier mode corresponding to the LSC. We have identified a threshold value of the rotation rate below which the flow is mainly LSC dominated. Further, we have computed the variation in the boundary layer width in the plane in which LSC is confined and quantitatively showed its overall effect on the heat transfer rate. We have also identified the presence of different kind of reorientations, like, rotation-led and cessation-led, same as RBC and have extensively showed the effect of rotation on these events. Interestingly through our numerical work we have been able to consider the statistics of reorientations, which have not been previously attempted numerically and obtain the power law characteristics of them which are very close to the earlier experimental results. 
\par
The paper is organized as follows. Section \ref{Sec:NUM_DET} details the mathematical formulations and numerical methods used along with grid independence study. The results of the present study with detailed discussions on the dynamics of reorientations of LSC and flow statistics are summarized in section \ref{Sec:Results}. Finally, the present work is concluded by section \ref{Sec:Conclusion} by outlining the main findings.
\section{Numerical details}\label{Sec:NUM_DET}
\begin{figure}[ht!]
\centering
\includegraphics[scale=0.30]{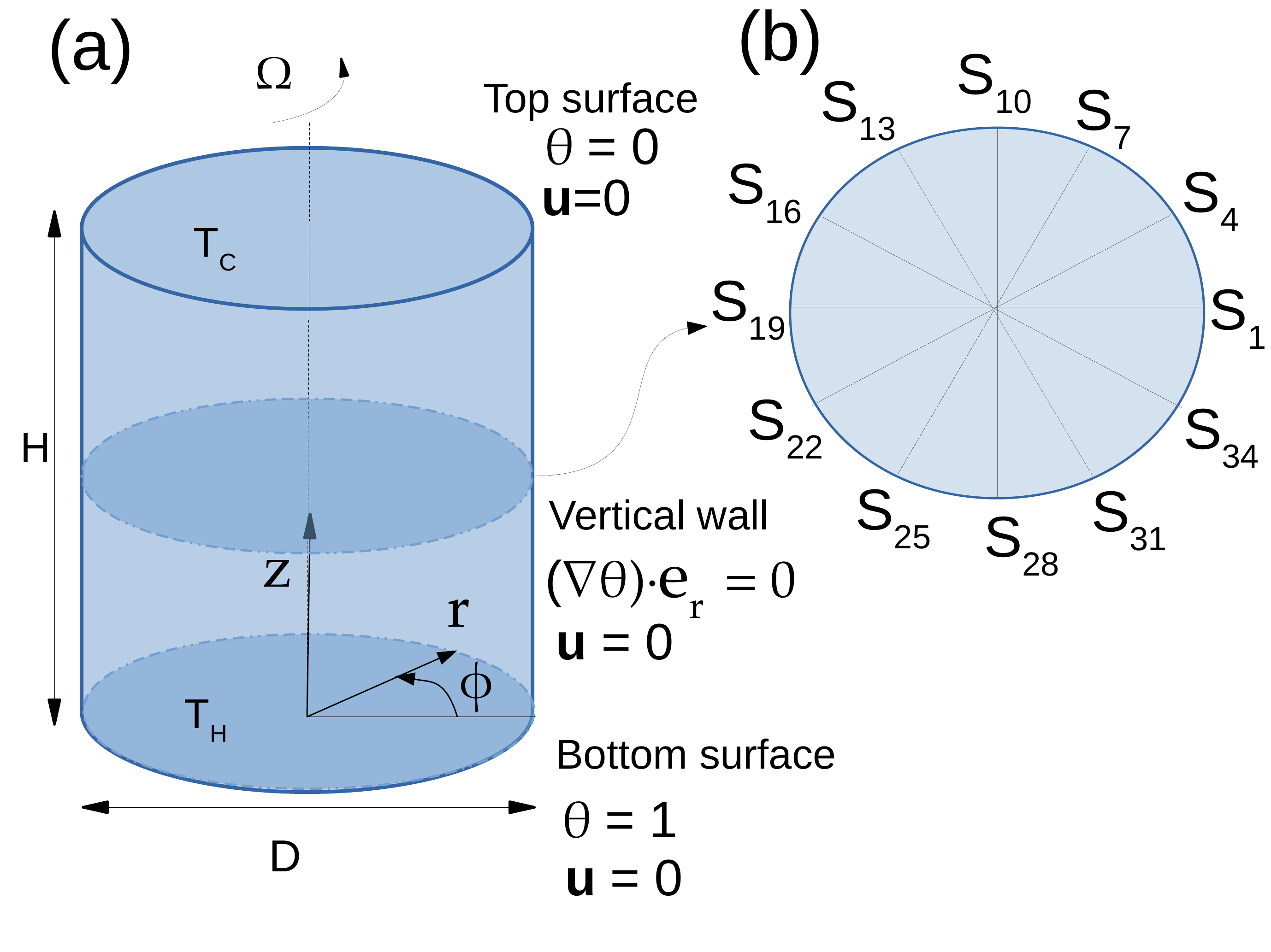}\
\caption{(a) Schematic diagram of the flow domain with relevant boundary conditions. (b) Schematic representation of the location of numerical probes at mid-vertical cross section of the cylinder used for signal analysis.}
\label{flowdomain}
\end{figure}

In this section, we briefly describe the mathematical formulations of the problem followed by grid independence study. A schematic diagram of the flow domain with relevant boundary conditions used is shown in Fig. \ref{flowdomain}(a). We consider a cylindrical cell of unit aspect ratio $\Gamma=D/H$ with bottom heated ($T_H$) and top cooled ($T_C$) configuration.
Rotation is applied about the vertical axis of the cylinder. Mathematically the problem can be described using the momentum, mass and energy conservation equations written in the rotating frame of reference using the Boussinesq approximation as
\small
\begin{eqnarray}
\frac{\partial \bm{u}}{\partial t}+\bm{u}\cdotp \nabla \bm{u} &=& -\nabla P + \sqrt{\frac{Pr}{Ra}} \nabla^{2} \bm{u}  +  \theta \bm{\hat{k}}-\frac{1}{Ro}(\bm{\hat{k}} \times \bm{u}) \label{gov_eqn1} \\
\nabla \cdotp \bm{u} &=& 0  \label{gov_eqn2}\\
\frac{\partial \theta}{\partial t} + \bm{u}\cdotp \nabla \theta &=&  \frac{1}{\sqrt{Pr Ra}} \nabla^{2}\theta  \label{gov_eqn3}
\end{eqnarray}
\normalsize
\smallskip
\noindent
where $\bm{u}$ is the velocity in the rotating frame, $P$ is the modified pressure, $\hat{k}$ is the unit vector in the vertical direction and $\theta$ ($=(T-T_{C})/\Delta T$) is the non-dimensional temperature. We use $H$, $V=\sqrt{g\beta \Delta T H }$ and $\Delta T = (T_H - T_C)$ as the scales for normalization of length, velocity and temperature, respectively. 
\smallskip
\par
The governing Eqs. \ref{gov_eqn1}- \ref{gov_eqn3} are solved using finite volume formulation with collocated arrangement of variables. 
No-slip boundary conditions are implemented on all the surfaces. For temperature, horizontal surfaces are iso-thermal and the lateral wall is adiabatic.
The convective term is approximated using the $2$nd-order Adams-Bashforth scheme while the buoyancy and diffusive terms are approximated by the Crank-Nicholson scheme. Time increments of $\Delta t=10^{-3}$ and $5\times10^{-4}$ are used for  $Ra=2\times10^6$ and $2\times10^7$, respectively. To adequately address the issue of velocity-pressure coupling, a predictor-corrector based momentum interpolation method \cite{Peter_De_Ocean} is used as the solution technique. Bi-Conjugate 
Gradient Stabilized (BiCGSTAB) technique preconditioned by Stone's Strongly Implicit Procedure (SIP) \cite{Ferziger_Peric} is used to solve all the resulting sparse 
linear systems. For details of the numerical procedures Peter and De \cite{Peter_De_Ocean} can be referred.
\begin{table*}[t!]
\caption{Different grids with corresponding number of grid points within the thermal boundary layer $N_{BL}$, the critical grid parameters $l_{max}/\eta_{k}$, $\Delta z_{max}/\eta_{k}$ and $\Delta_{max}/\eta_{k}$ followed by volume averaged Nusselt number $Nu_{vol}$, Nusselt number at the bottom wall $ Nu$ and ratio of numerical to analytical dissipation rates ($\langle\bigepsilon^{num}_u\rangle / \langle\bigepsilon^{th}_u\rangle $ and $\langle\bigepsilon^{num}_{\theta}\rangle / \langle\bigepsilon^{th}_{\theta}\rangle$) at $Ra=2\times10^6$ and $2\times10^7$ for non-rotating and high rotation case. }
\centering
\begin{tabular}{cc cc cc  cc cc cc cc cc cc }
\hline \hline 
\vspace {0.5em} 
$Ra$&$Ro^{-1}$ &$N_{r} \times N_{\phi} \times N_{z}~ $&$N_{BL}$&\large{$\frac{l_{max}}{\eta_{k}}$}&\large{$\frac{\Delta z_{max}}{\eta_{k}}$}&\large{$\frac{\Delta_{max}}{\eta_{k}}$}&$~ Nu_{vol}~$&$~Nu~$& \large{$~\frac{\langle\bigepsilon^{num}_u\rangle}{\langle\bigepsilon^{th}_u\rangle }~$}&\large{$~\frac{\langle\bigepsilon^{num}_{\theta}\rangle}{\langle\bigepsilon^{th}_{\theta}\rangle }~$}&\\[.10ex] 
\hline 
  \hline
 \small \multirow{2}{*}{$2\times10^6$}& 0 & $43\times 101 \times 85$&$9$&$0.804$ & $0.515$&$0.455$&$10.75$&$10.71$&$0.981$& $0.987$&\\
\small & 10 & $43\times 101 \times 85$&$13$&$0.715$ & $0.457$&$0.404$&$3.68$&$3.69$&$0.978$& $0.990$&\\
\hline
 \small \multirow{2}{*}{$2\times10^6$}& 0&$50\times 101 \times 101$&$11$&$0.804$ & $0.432$&$0.409$  &$10.68$&$10.66$&$0.990$& $0.983$&\\
 \small & 10&$50\times 101 \times 101$&$15$&$0.715$ & $0.384$&$0.363$  &$3.68$&$3.69$& $0.990$& $0.982$&\\
\hline
 \small \multirow{2}{*}{$2\times10^6$}& 0&$60\times 101 \times 115$&$13$&$0.804$ & $0.380$&$0.367$  &$10.65$&$10.60$&$0.991$& $0.984$&\\
 \small & 10&$60\times 101 \times 115$&$19$&$0.715$ & $0.338$&$0.326$  &$3.67$&$3.68$&$0.985$&$0.984$&\\
 \hline
 \hline
 \small \multirow{ 2}{*}{ $2\times10^7$}& 0 &$93\times 175 \times 185$&$12$&$0.968$& $0.492$&$0.492$     &$20.47$&$20.42$&$0.978$&$0.986$&\\
 \small & 10 &$93\times 175 \times 185$&$27$&$0.724$& $0.368$&$0.350$    &$6.49$&$6.50$&$0.984$&$0.989$&\\
  \hline
\small\multirow{ 2}{*}{ $2\times10^7$}& 0&$101\times 175 \times 201$&$13$&$0.968$& $0.453$&$0.453$     &$20.46$&$20.46$&$0.980$&$0.988$&\\
\small & 10&$101\times 175 \times 201$&$30$&$0.724$& $0.339$&$0.331$     &$6.37$&$6.38$&$0.986$&$0.992$&\\
  \hline
\small \multirow{ 2}{*}{ $2\times10^7$}& 0 &$115\times 175 \times 229$&$15$&$0.968$& $0.391$&$0.391$     &$20.42$&$20.39$&$0.982$&$0.988$&\\
\small & 10 &$115\times 175 \times 229$&$32$&$0.724$& $0.293$&$0.304$     &$6.39$&$6.40$&$0.989$&$0.993$&\\
\hline \hline
\end{tabular}
\label{tab:grid_stats_full}
\smallskip
\end{table*}
\smallskip
\par
While reporting direct numerical simulations an important requirement is that the grid is resolved adequately to capture the flow physics correctly at the same time it consumes less computational resources.
Two criteria are imposed for calculating the grid sizes. Firstly, the boundary layers are aptly resolved, for which we necessitate that at least 10 grid points are present within the boundary layers (both near horizontal and lateral walls). As a result $\Delta z_{min} \approx \delta_{\theta} /10 $ and $\Delta z_{min} \approx \Delta r_{min}$. Secondly, the  mesh size must be of the order less than or equal to the Kolmogorov scale. We implement the conditions $\Delta_{max}\lesssim  \eta_K$,  $l_{max}(=2\pi r/N\theta)  \lesssim \eta_K$ , $\Delta z_{max} < \eta_K$ and $\Delta r_{max} \approx \Delta z_{max}$. Here $\Delta_{max}$ indicates the maximum of mean grid size calculated as $(r\Delta \phi \Delta r \Delta z)^{1/3}$ and $\eta_K$ is the Kolmogorov scale estimated using the analytical formula $\eta_K \approx \pi (Pr^2/RaNu)^{1/4}$ given by Gr\"{o}tzbach \cite{grotzbach}. Table \ref{tab:grid_stats_full} shows the different computational grids with corresponding number of grid points within the boundary layer followed by the maximum grid size in the azimuthal and axial direction, and the maximum mean grid size, normalized with the Kolmogorov scale for both the Rayleigh numbers at $Ro^{-1}=0$ and $10$.

\par
The Nusselt number at the bottom wall is computed as $Nu = \langle \partial \theta / \partial z \rangle_{A,t}$, where $\langle .. \rangle_{A,t}$ represents averaging over horizontal plane and time. Further, the volume averaged Nusselt number over the entire domain $ Nu_{vol}=1+\sqrt{RaPr} \langle w\theta \rangle$ ($\langle ..\rangle$ denotes averaging over volume and time) is also calculated. Both values agree well and the difference between the $ Nu_{vol}$ (or $Nu$ ) values computed at the coarsest and finest grid for both $Ra$ is less than $2\%$, which indicates a good grid convergence. To further confirm the spacial resolution, the ratio of the numerical to the theoretical estimate of the  viscous ($\langle\bigepsilon^{num}_u\rangle/\langle\bigepsilon^{th}_u\rangle$) and thermal dissipation rates ($\langle\bigepsilon^{num}_{\theta}\rangle/\langle\bigepsilon^{th}_{\theta}\rangle$) are calculated for both non-rotating ($Ro^{-1}=0$) and sufficiently high rotation rate $Ro^{-1}=10$.
Here $\langle\bigepsilon^{num}_u\rangle =\nu V^{2}H^{-2}\langle\lvert\nabla \bm{u}\rvert^{2}\rangle$~and~$\langle\bigepsilon^{num}_{\theta}\rangle=\alpha(\Delta T)^{2}H^{-2}\langle\lvert\nabla\theta \rvert^{2}\rangle$~are the numerically calculated values of viscous and thermal dissipation rates, respectively, while $\langle\bigepsilon^{th}_u\rangle=\nu^{3}(Nu-1)RaPr^{-2}H^{-4} $ and $\langle\bigepsilon^{th}_{\theta}\rangle =\alpha (\Delta T)^{2}NuH^{-2}$ are their analytical counterparts\cite{Shraiman_Siggia_1990}.
%
Note that the theoretical estimates of the dissipation rates are functions of Nusselt number, for which we substitute the corresponding numerically obtained value \cite{stevens5,MishraDe}.
The ratios are always close to unity (minimum $\approx 0.978$), suggesting that the present grid is well resolved to perform the proposed numerical calculations.
Considering the computational cost and the accuracy of the above mentioned parameters, a grid size of $50\times 101 \times 101$ and  $101\times 175 \times 201$ are chosen for $Ra=2\times10^6$ and $2\times10^7$, respectively. 
\vspace{3 mm}
\section{Results and Discussions}\label{Sec:Results}
In this section we discuss the principal findings of the present work.
First we establish the presence of LSC in our simulations of RBC with and without rotation. 
Then we study the characteristics of LSC and its relation with heat transfer, boundary layer dynamics, and the thermal and viscous dissipation rates. 
This is followed by a detailed discussion on the dynamics of LSC, quantified by the Fourier mode analysis. Finally, we investigate the statistics of reorientations of LSC.

\subsection{Identification of LSC}
Large-scale circulation is a coherent structure in RBC where the thermal plumes organize to form a ``large-scale'' convection roll with hot and cold fluid rising and dipping, respectively, along opposite sides of the lateral wall. 
Over the years a number of experimental \cite{Krishnamurthi_Howard_1981,Funfschilling_Ahlers_2004,Sun_etal_2005a,Brown_Ahlers_JFM} and numerical \cite{Stringano_Verzicco_JFM_2006,Verzicco_Camussi_1999,MishraDe} studies have reported presence of LSC and its associated dynamics for non-rotating RBC. However, such attempts on the dynamics of LSC for flows modulated by rotation are few. Here we focus on the identification of LSC in rotating and non-rotating RBC using flow structures and time signals of vertical velocity.   

\par
\begin{figure}
\centering
\includegraphics[scale=0.110]{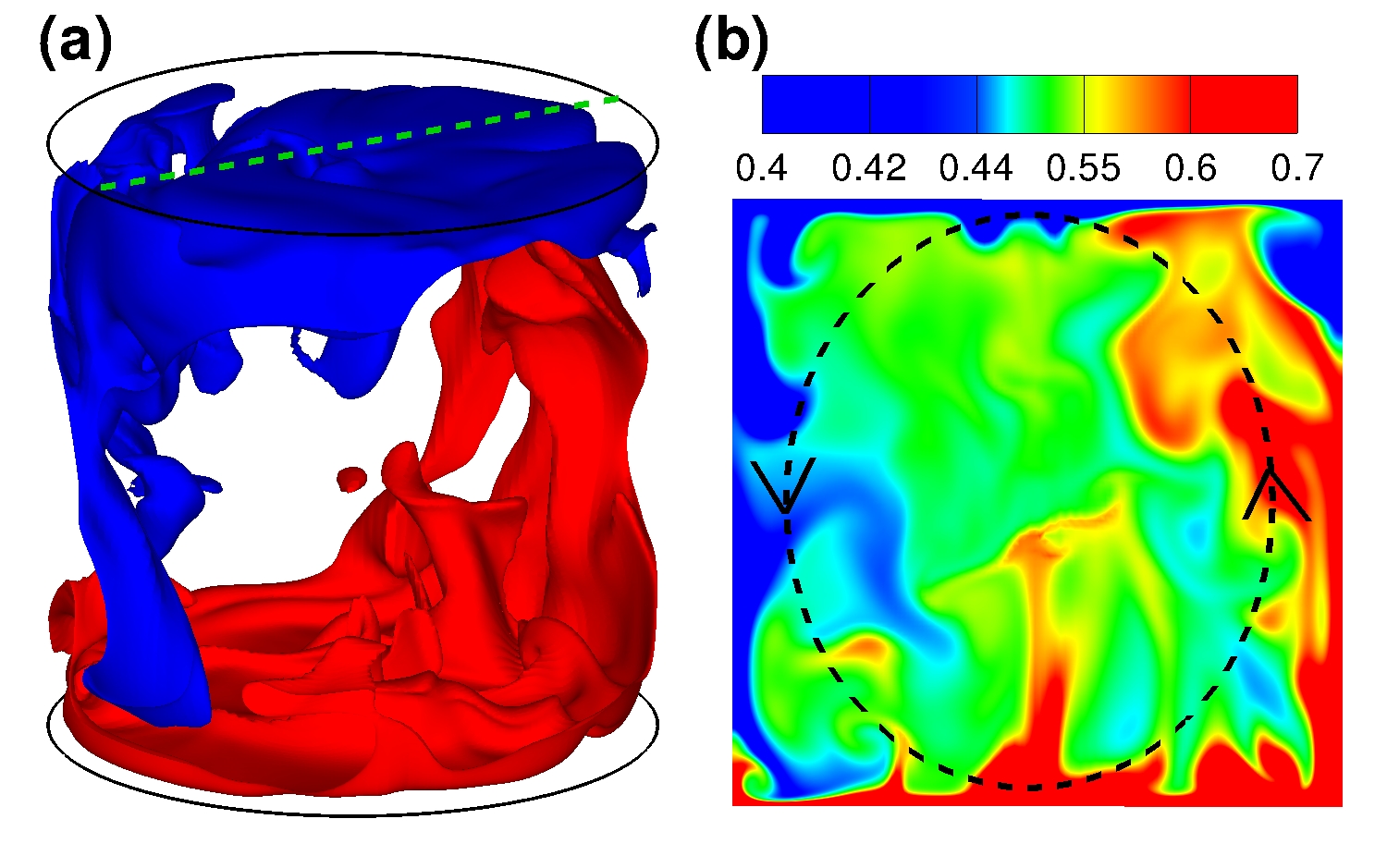}\\
\includegraphics[scale=0.110]{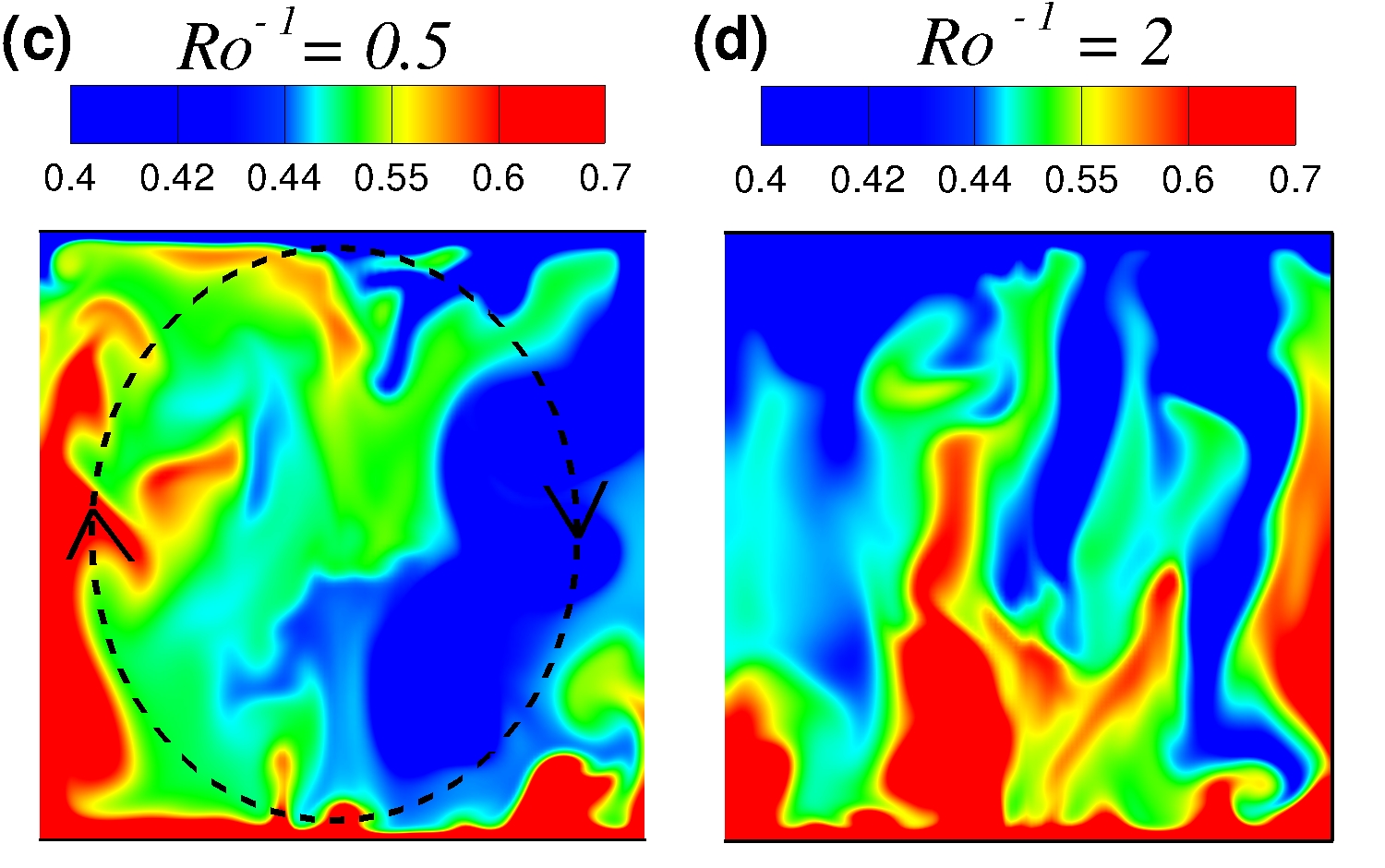}
\caption{(Colour online) (a) Temperature iso-surfaces at $\theta=0.4$ (blue) and $\theta =0.6$ (red) indicating LSC for $Ra=2\times10^7$  and $Ro^{-1}=0.2$. The dotted line indicates the azimuthal orientation of LSC. (b) Temperature contour at the vertical plane where the LSC is oriented. The dotted line indicates the direction of circulation. (c) Temperature contour at LSC plane for $Ro^{-1}=0.5$. (d) Temperature contour at $\phi=0$ plane for $Ro^{-1}=2$.}
\label{LSC_Diss1}
\end{figure}

An instantaneous snapshot of the temperature iso-surfaces is shown in Fig. \ref{LSC_Diss1}(a) for $Ra=2\times10^7$  and $Ro^{-1}=0.2$.   Xi \emph{et al.} \cite{Xi_Lam_Xia_2004} experimentally studied RBC and observed that LSC is a result of organization of the thermal plumes. On a similar note, we observe that hot and cold plumes are clustered separately along opposite side of the cylinder, thus indicating an LSC structure. The dotted line indicates an approximate azimuthal orientation of LSC.
The temperature contours at the vertical mid-plane along which LSC is aligned [indicated in Fig. \ref{LSC_Diss1}(a)] is shown in  Fig. \ref{LSC_Diss1}(b). For clarity, we have shown a dashed line which gives the directional overview of LSC in that plane.  The hot (cold) plumes rise (fall) from the bottom (top) surface,  proceed along the cylinder bulk and plunge into the boundary layer near the opposite wall. A similar flow structure is observed at higher rotation rate  ($Ro^{-1}=0.5$) shown in Fig. \ref{LSC_Diss1}(c), with a change in the sense of circulation. We observe similar large-scale structure at $Ro^{-1}<1$.
As we further increase the rotation, the coherence at large scale is lost. For example at $Ro^{-1}=2$ shown in in Fig. \ref{LSC_Diss1}(d), the temperature contours indicate a distorted structure.

\par
\begin{figure}[h!]
\centering
\includegraphics[scale=1.150]{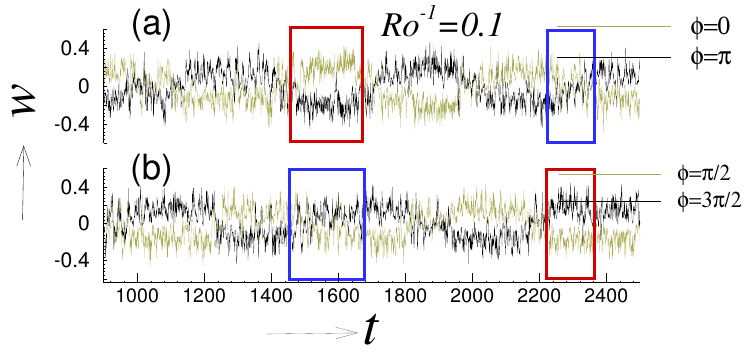}
\caption{(Colour online) Time traces of vertical velocity for $Ra=2\times10^7$ and $Ro^{-1}=0.1$. The red and blue boxed regions indicate the anti-correlated signals and high-variance data, respectively, at diameters which are $\pi/2$ apart.}
\label{Ra7_Sig2}
\end{figure}
\begin{figure}[h!]
\centering
\includegraphics[scale=1.30]{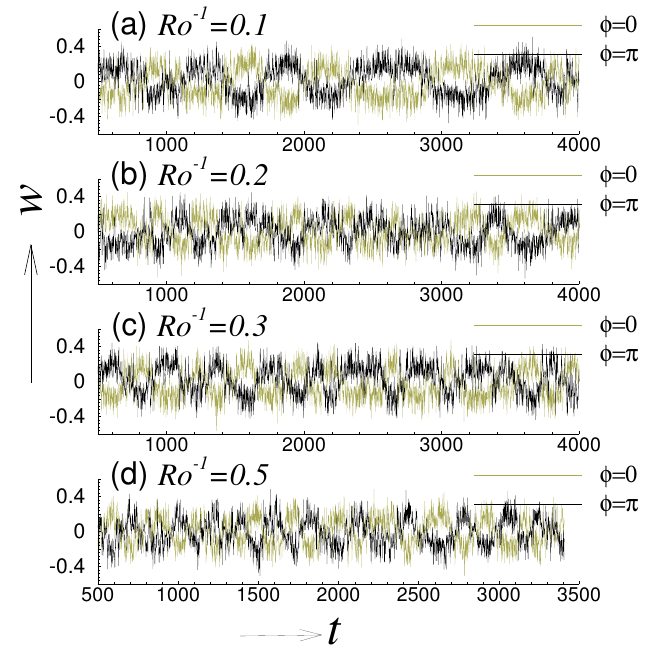}
\caption{Time traces of vertical velocity  from numerical probes which are $\pi$ apart in the azimuthal direction at $Ra=2\times10^7$ for different $Ro^{-1}$. The switching frequency of the signals increases with the increase in rotation rate.}
\label{Ra7_Sig1}
\end{figure}
\begin{figure}
\centering
\includegraphics[scale=0.55]{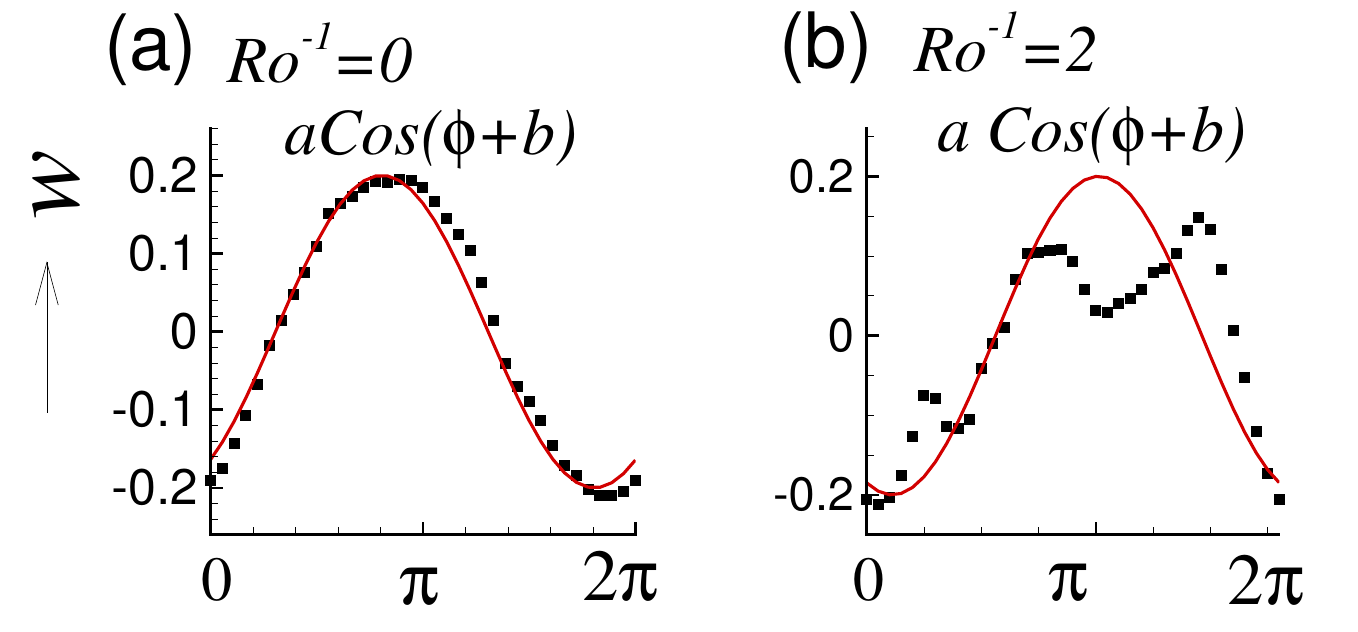}\\
\includegraphics[scale=0.55]{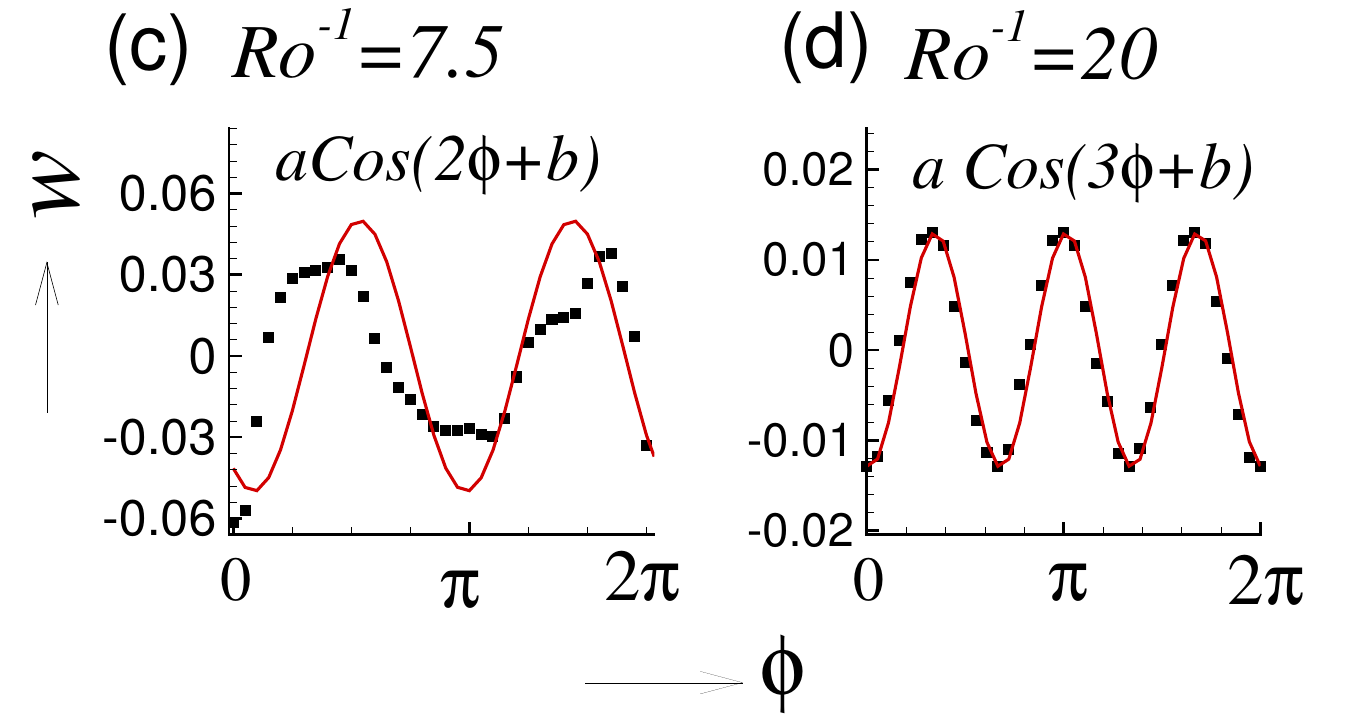}
\caption{The temporal averaged vertical velocity profile along the azimuthal direction for $Ra=2\times10^6$ at (a)$Ro^{-1}=0$, (b)$Ro^{-1}=2$, (c)$Ro^{-1}=7.5$ and (d)$Ro^{-1}=20$. The solid red line indicates a pure cosine fit.}
\label{fig:Wazi}
\end{figure}

In order to characterize the LSC, time signals of vertical velocity are recorded at $36$ equispaced azimuthal stations that are located at the mid-vertical plane ($z=0.5$) near the lateral wall ($r=0.4R$).
Figure \ref{Ra7_Sig2} shows time signals of vertical velocity at either ends of two diametrical planes separated by an angle $\pi/2$. The LSC is primarily identified by the occurrence of non-zero mean vertical velocities that are anti-correlated, shown in rectangular windows (red). Note while the LSC persists along a diameter, high-variance fluctuations are observed on the other and they switch between each other. 
The time over which the LSC persists along a diameter depends strongly on the rotation rate. Figure \ref{Ra7_Sig1} shows that as rotation rate is increased the switching frequency of the signals increases. These are partly associated with reorientations of LSC along the azimuthal plane, discussed in subsequent sections. 
The time average vertical velocity along the above-mentioned azimuthally equispaced numerical probes is shown in Fig. \ref{fig:Wazi}, where the averaging is carried out within a time over which LSC remains in a particular plane. For the non-rotating case (and low rotation rates) the profile shows a cosine function with a single cycle spanning the entire ($2\pi$) domain [see Fig. \ref{fig:Wazi}(a)] which confirms the presence of LSC structure, and is consistent with previous studies \cite{Cioni_etal_1997,StevensPoF2011,Xi_etal_PRL_2009,Brown_Ahlers_JFM}. This particular nature of the flow in the azimuthal direction can be associated with the dipolar structure.
As rotation rate increases to $Ro^{-1}=2$ we find a distortion from the cosine fit, as shown in Fig. \ref{fig:Wazi}(b).
Further increase in rotation leads to deviation from the dipolar structure and flow appears to be more close to the quadrupolar structure, indicated by  two pairs of $+$ve and $-$ve peaks in the velocity profile [refer Fig. \ref{fig:Wazi}(c)].
At very high rotation rates ($Ro^{-1}=20$), the vertical velocity profile appears to be more close to a sextupolar behaviour with $+$ve (up-flow) and $-$ve (down-flow) velocity separated azimuthally by $\pi/3$. Note that at lower and higher rotation rates we observe almost perfect cosine fits for the vertical velocity profile, while at intermediate rotation rates it deviates from cosine fit.


\par
\begin{figure}
\centering
\includegraphics[scale=0.40]{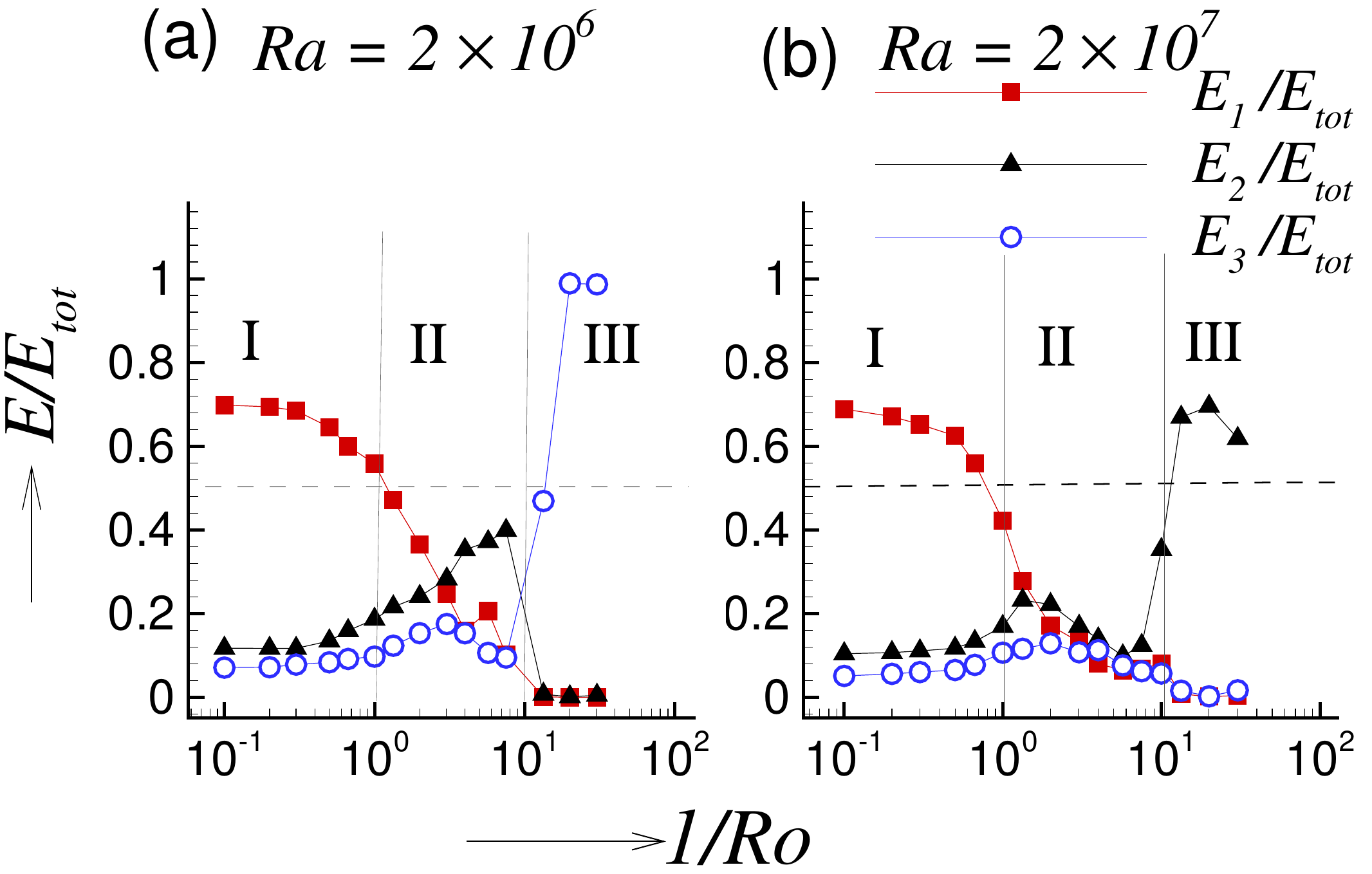}
\caption{(Colour online) The variation of energy fraction of the first three Fourier modes with rotation rate for $Ra=2\times10^6$ and $2\times10^7$.}
\label{fig:Fmodes}
\end{figure}

Following the works of  Cioni \emph{et al} \cite{Cioni_etal_1997}, Brown and Ahlers \cite{Brown_Ahlers_JFM} and Mishra \emph{et al.}\cite{MishraDe},
to get further insights into LSC and its dynamics, we carry out Fourier transform of the vertical velocity recorded along the azimuthal direction at mid-vertical plane.
Note that $u_j$ is the velocity signal from $N~(=36)$ data points along the cylinder azimuth [refer Fig. \ref{flowdomain}(b)] and its Fourier transform is written as
\begin{eqnarray}
 \hat{u}_{k}= \sum_{j=1}^{N} u_{j}  e^{-i2\pi k j/N} \label{gov_eqn4}
\end{eqnarray}
\noindent
where $\hat{u}_{k}$ represents the $k-$th Fourier mode with $\Phi_k$ being its phase.
The first ($\hat{u}_{1}$), second and third modes are associated with the dipolar, quadrupolar and sextupolar structure of the flow, respectively \cite{Xi_Zhang_JFM_2016}. 
Based on the formalism proposed by Kunnen \emph{et al.} \cite{kunnen2}, Weiss and Ahlers \cite{Weiss_Ahlers_JFM_2011_LSF} and   Xi \emph{et al.} \cite{Xi_Zhang_JFM_2016}, we have characterized the LSC using the fraction of energy contained in the $k-$th mode $E_{k}/E_{tot}$, where $E_{k}=|\hat{u}_k|^2$ and $E_{tot}= \sum_{k=1}^{N/2} E_{k}$. The evidence of LSC is often ascertained from the energy fraction of first Fourier mode ($E_{1}/E_{tot}$).
Figure \ref{fig:Fmodes} shows the variation of the energy fraction of first three Fourier modes with rotation rate.
We identify different flow regimes based on the dominance of specific modes. In regime \Romannum{1} ($0 \leq Ro^{-1} \lesssim 1$), the first Fourier mode is clearly dominant with $E_{1}/E_{tot} > 0.5$ and is identified as the LSC regime. Consequently, the average vertical velocity profile along the azimuthal direction clearly shows a perfect cosine fit [see Fig. \ref{fig:Wazi}(a)] as discussed before. A decrease in the energy fraction of the first Fourier mode accompanied by increase in that of second ($E_{2}/E_{tot}$) with the increase in rotation rate is observed here. At high rotation rates regime \Romannum{2} is identified ($1 \lesssim Ro^{-1} \lesssim 10$) where all the three modes share almost equal energy. As a result the vertical velocity profiles in this regime are far from perfect cosine fit, as shown in Figs. \ref{fig:Wazi}(b) and (c).
For $Ra=2\times10^6$ although the second mode marginally dominates at moderately high rotation rates ($5 < Ro^{-1} \lesssim 10$), it declines at extremely high rotation rates. Here, we even observe a third Fourier mode dominated flow for $Ro^{-1} \gtrsim 10$. This is identified as regime \Romannum{3}, where the higher mode (second or third) shares more than $50\%$ of the total energy.
For $Ra=2\times10^7$ the second Fourier mode dominates over the other modes in  regime \Romannum{3}.


\par
\begin{figure}
\centering
\includegraphics[scale=0.290]{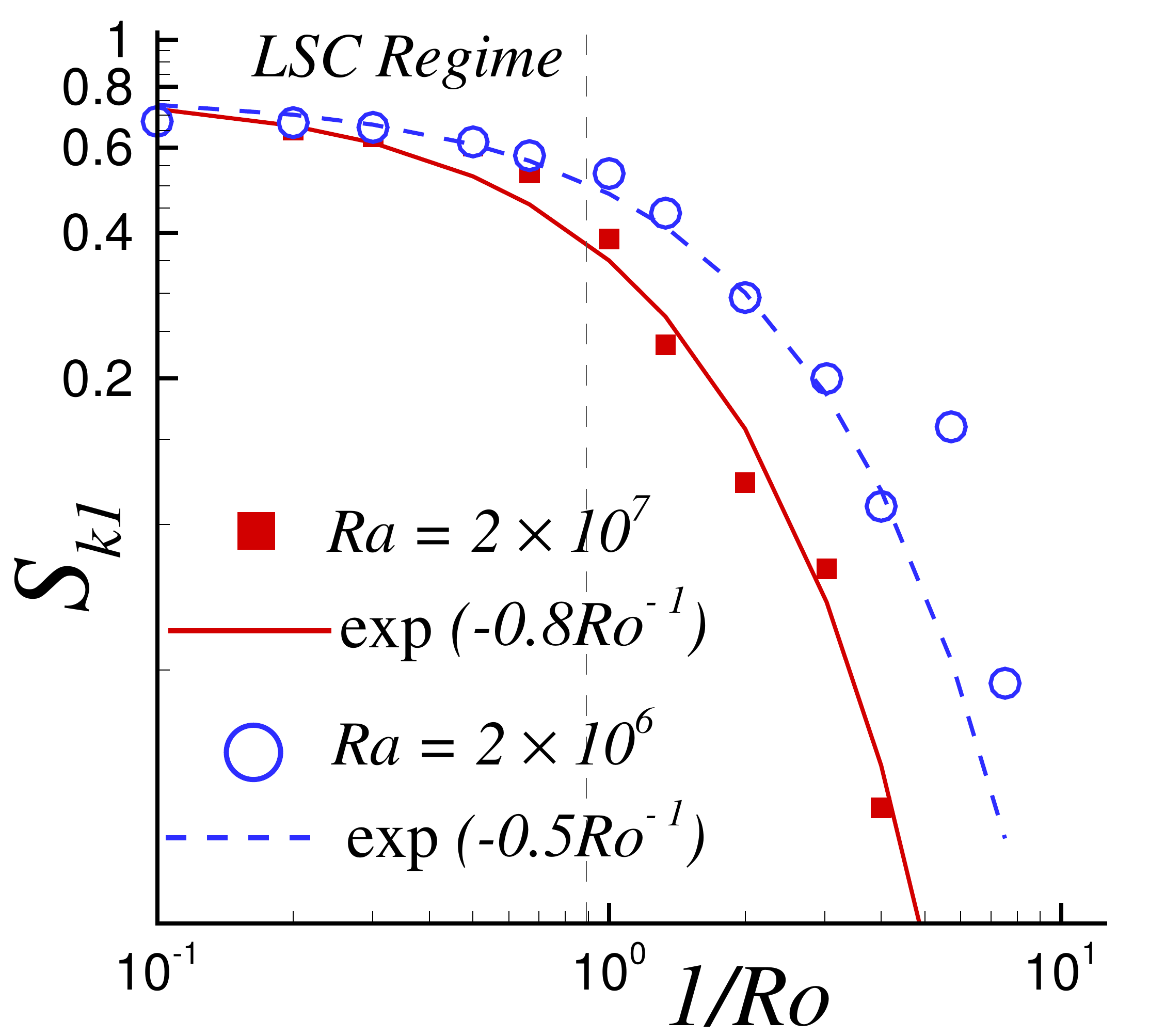}
\caption{Exponential decay of the strength of the first Fourier mode with the increase in rotation rate for $Ra=2\times10^6$ and $2\times10^7$.}
\label{fig:Sk}
\end{figure}

In order to make it more quantitative we also compute the strength of LSC ${S}_{k1}$ \cite{StevensPoF2011}, defined as

\begin{eqnarray}
{S}_{k1} = Max \Big[  \Big( \frac{ E_{1}}{ E_{tot}} -\frac{1}{N} \Big) / \Big(1-\frac{1}{N}\Big), 0      \Big]
\end{eqnarray}
\noindent
The value of ${S}_{k1}$ lies between $0$ and $1$. The strength close to unity indicates that most of the energy is contained in the first Fourier mode, i.e., $E_{1}/E_{tot}\approx1$ and the azimuthal vertical velocity profile is close to a cosine fit, thus indicating LSC. On the other hand,  its value near to zero implies that the energy fraction of first Fourier mode is smaller ($E_{1}/E_{tot}<< 1$), in effect  most of the energy is contained in the higher modes and the flow behaviour is far from LSC structure.
In Fig. \ref{fig:Sk} we plot the circulation strength against the rotation rate. At low rotation rates the strength is as high as $0.7$, while it drops exponentially with the increase in rotation rate. We observe a perfect exponential fit as $S_{k1}\propto\exp (-0.5 Ro^{-1} )$ and $S_{k1}\propto\exp (-0.8 Ro^{-1} )$ for $Ra=2\times10^6$ and $2\times10^7$, respectively. The decay of $S_{k1}$ is at a higher rate for the higher Rayleigh number. Similar observation is made from the energy fraction of the first Fourier mode (refer Fig. \ref{fig:Fmodes}), where it declines steeply for $Ra=2\times10^7$.
We identify the LSC regime as $Ro^{-1}\lesssim1$ where the circulation strength is greater than $0.5$.
This is consistent with the observations made by Kunnen \emph{et al.} \cite{kunnen2} and Stevens \emph{et al.}\cite{stevens7}, where they reported the break-down of LSC for  $Ro^{-1}\gtrsim 0.86$.

\subsection{Effect of LSC on heat transfer and boundary layer dynamics}
In the previous subsection we identified several flow regimes based on the energy contained in the Fourier modes related to the large scale circulation. In this subsection we investigate the nature of heat transfer, dissipation rates and dynamics of boundary layer in those regimes. In addition we also explore 
how the presence of LSC overall affects the features of the boundary layer.

\begin{figure}
\centering
\includegraphics[scale=0.2750]{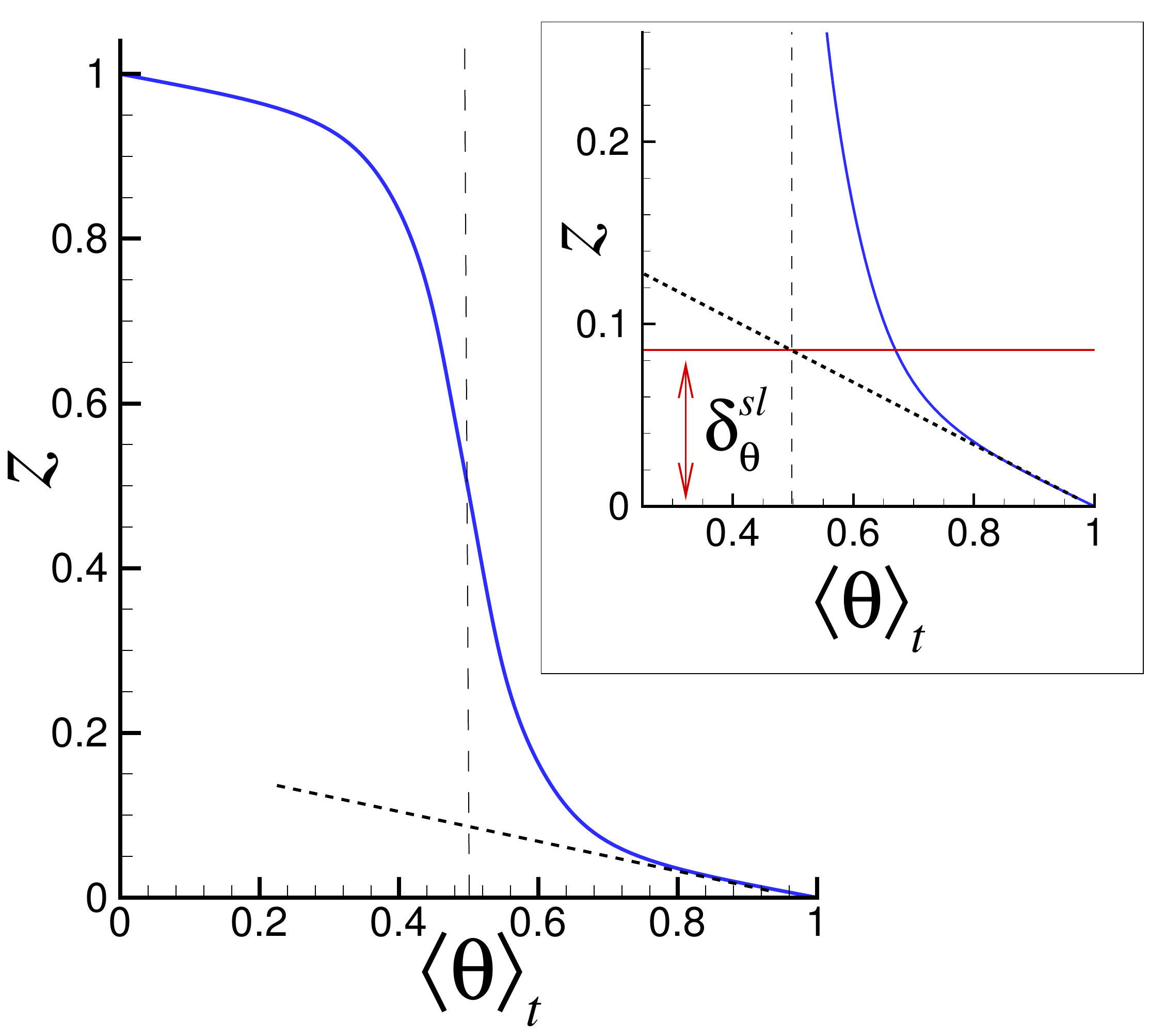}
\caption{(a) Schematic representation of the time averaged temperature profile showing the intersection of linear fit near the bottom plate with the bulk temperature. Inset shows the geometric construction of the thermal boundary layer thickness.}
\label{BLT_plot1}
\end{figure}

\begin{figure}
\centering
\includegraphics[scale=0.1300]{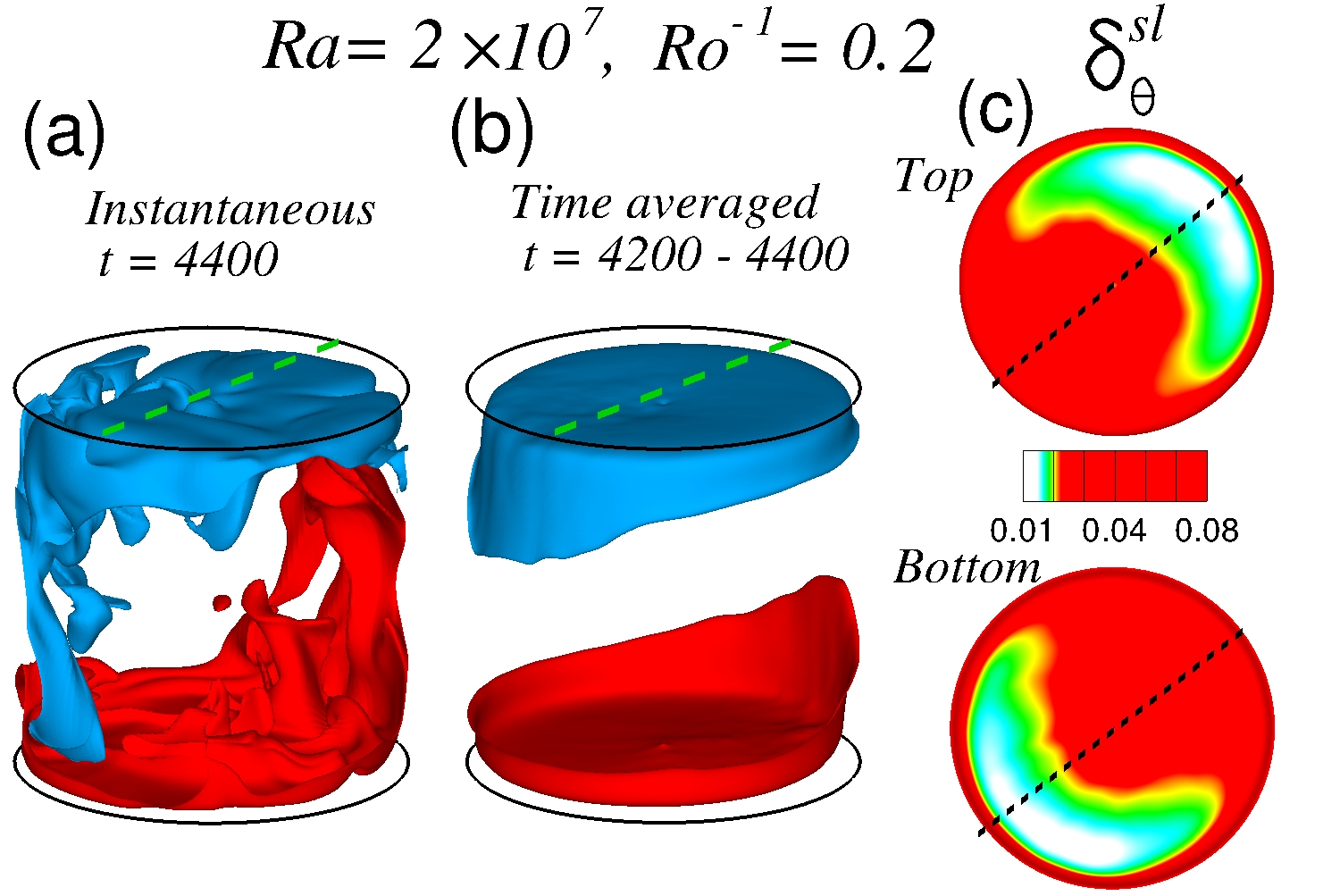}\\
\includegraphics[scale=0.330]{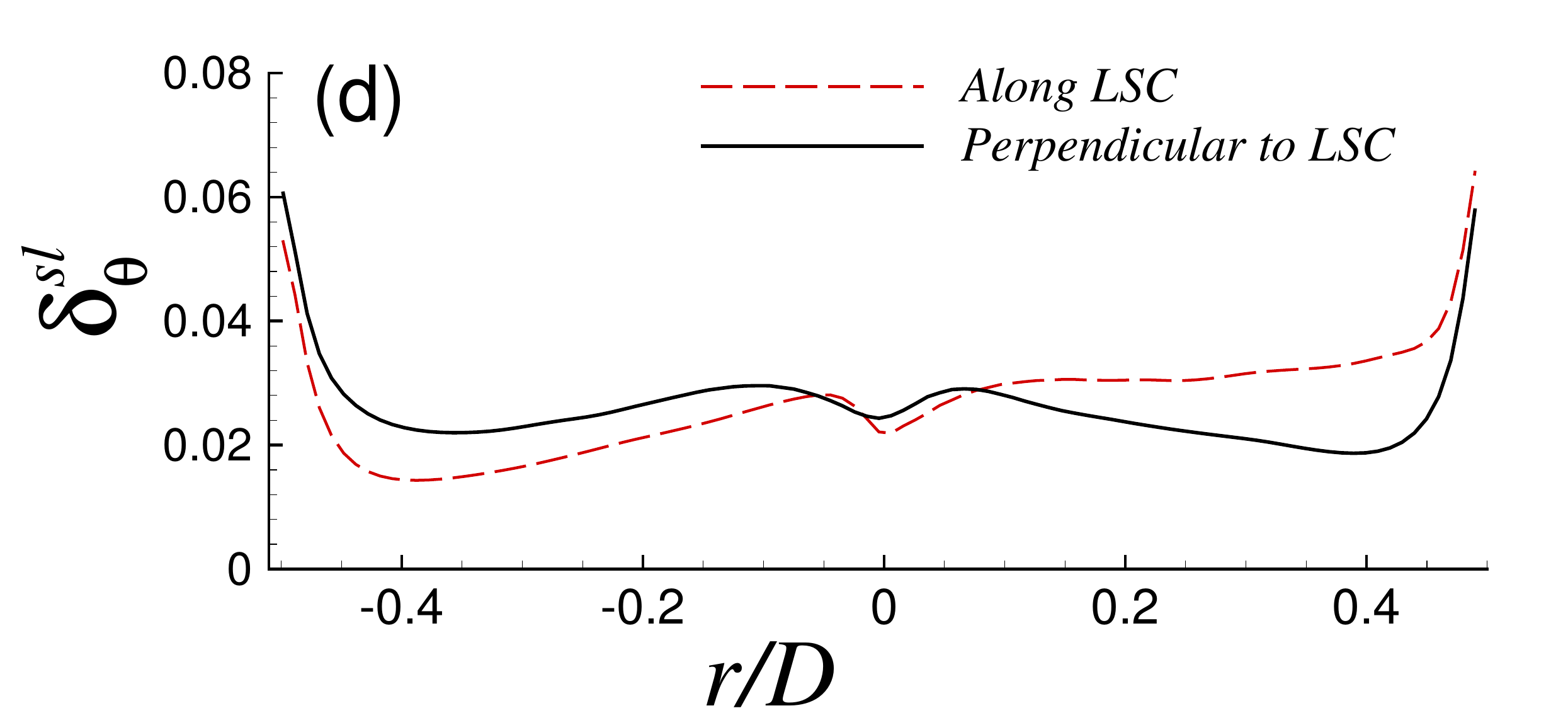}
\caption{(Colour online) (a) Instantaneous ($t=4400$) temperature iso-surfaces at $\theta=0.4$ (blue) and $\theta =0.6$ (red) indicating LSC. The dotted line represents the azimuthal orientation of LSC. (b) Time averaged temperature iso-surfaces at $\theta=0.4$ and $\theta =0.6$, obtained from a time span ($4200-4400$) at which LSC persists. (c) Contours of thermal boundary layer thickness near the top and bottom plate. (d) Variation of boundary layer thickness along and perpendicular to the direction of LSC, at the bottom. }

\label{BLT_plot_LSC}
\end{figure}

\begin{figure*}
\centering
\includegraphics[scale=0.250]{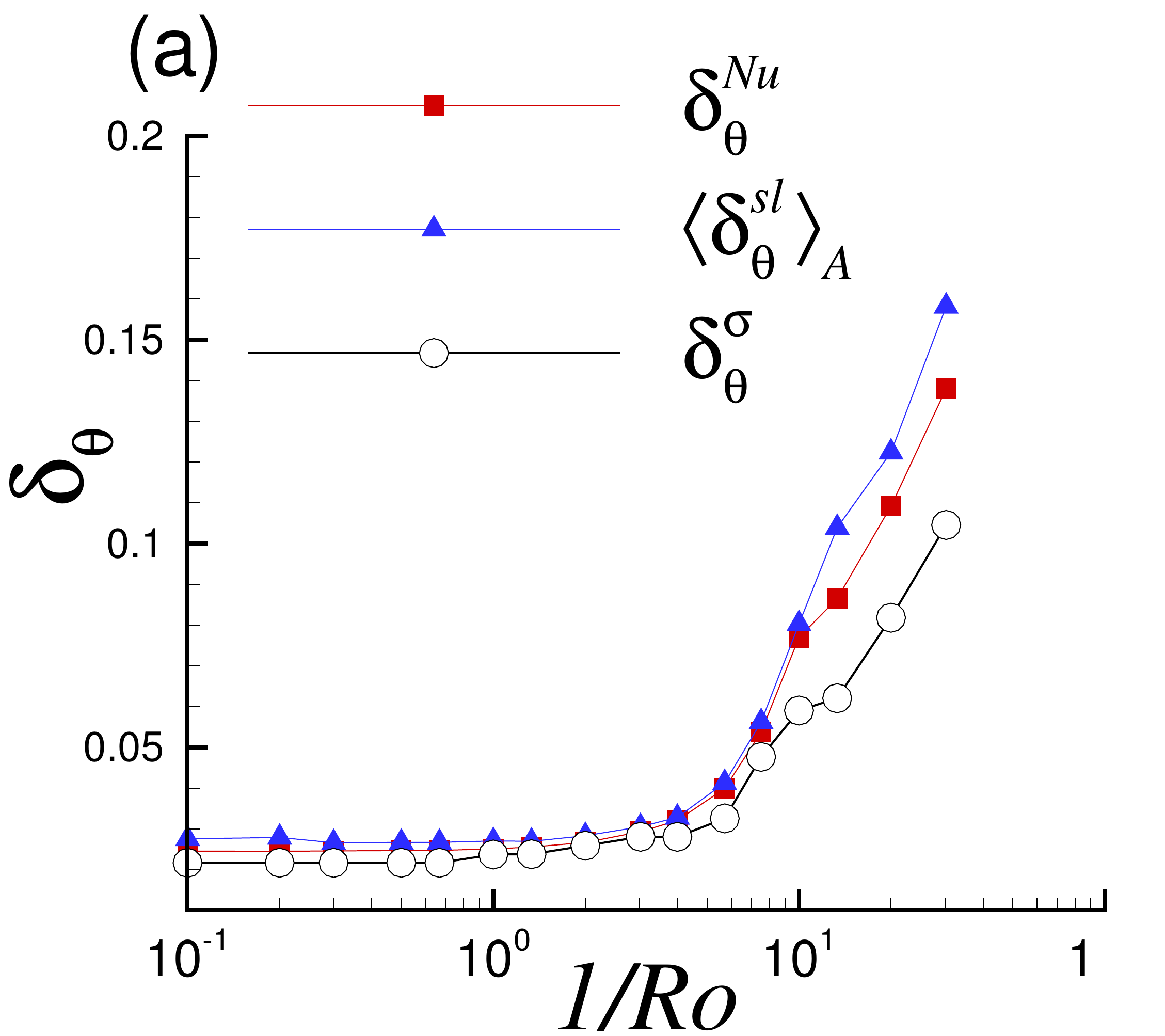}\includegraphics[scale=0.270]{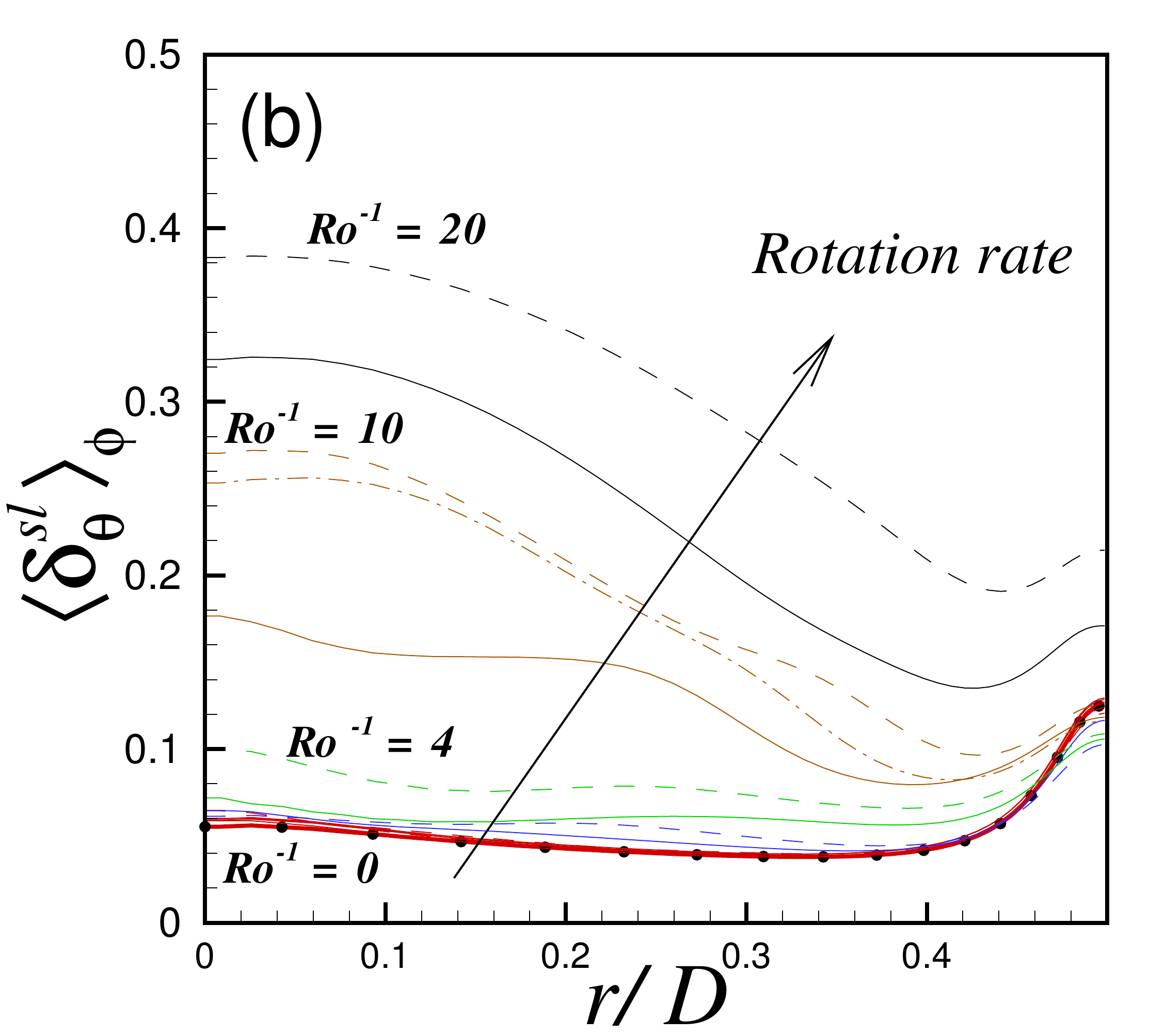}\includegraphics[scale=0.270]{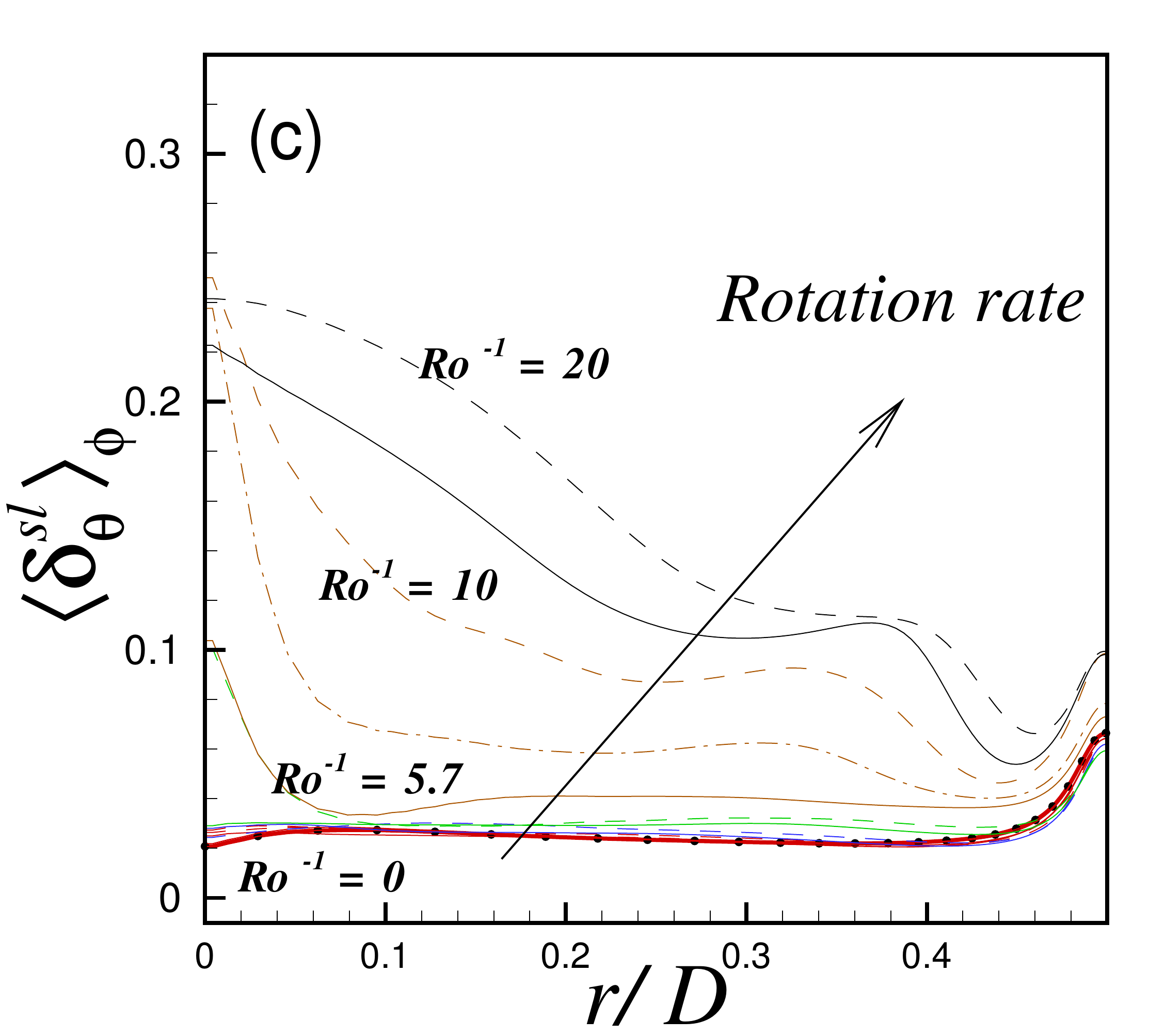}
\caption{(Colour online) (a) Thermal boundary layer thickness computed by different methods, as a function of inverse $Ro$ for $Ra=2\times10^7$. $\langle \delta_{\theta}^{sl} \rangle_A$ indicates the area averaged boundary layer thickness calculated from the slope method. (b) and (c) Variation azimuthal averaged thermal boundary layer thickness along the radial direction for $Ra=2\times10^6$ and $2\times10^7$, respectively, at different rotation rates.}
\label{Bl}
\end{figure*}

\par
In order to characterize the boundary layer we compute the thermal boundary layer thickness using three techniques. Firstly, $\delta_{\theta}^{Nu}$ which indicates the analytical thermal boundary layer thickness obtained from the relation $\delta_{\theta}^{Nu}  \approx 1/2  Nu $ \cite{Belmonte_Tilgner_Libchaber,Verzicco_Camussi_1999}. 
Secondly, $ \delta_{\theta}^{\sigma}$  which corresponds to the  boundary layer thickness computed as the vertical distance of the peak rms of temperature from the bottom plate \cite{Verzicco_Camussi_2003}. 
Finally, $ \delta_{\theta}^{sl}$ which is the boundary layer thickness calculated as the vertical distance of the point from the bottom plate where the linear fit of the time-averaged temperature profile approaches the bulk temperature \cite{Zhou_n5_PoF_2011,Lui_Xia}. A sample temperature profile is shown in Fig. \ref{BLT_plot1} where the geometric construction of the thermal boundary layer thickness is shown in the inset. Similar technique is adopted to compute the boundary layer thickness for all the spatial locations ($r,\phi$).

\par
At this point we are interested to see how the boundary layer thickness varies along and perpendicular to the plane containing LSC.
Figure \ref{BLT_plot_LSC}(a) shows the instantaneous temperature iso-surfaces indicating LSC structure. The time averaged flow structure for the same is shown in Fig. \ref{BLT_plot_LSC}(b), which is used to compute the thermal boundary layer thickness. Here averaging is performed within a time-span at which LSC persists in a particular plane. 
The planar contours thus obtained near the top and bottom plates are shown in Fig. \ref{BLT_plot_LSC}(c), where the orientation of LSC is shown by dashed line.
Close to the bottom plate, thermal boundary layer is thicker near the region where hot plumes rise and is thinner near the opposite side of the lateral wall where the cold plumes plunge into the boundary layer. Similar observation is made near the top plate. Further, we plot $\delta_{\theta}^{sl}$ along and perpendicular to the direction of LSC in Fig. \ref{BLT_plot_LSC}(d). Note that perpendicular to LSC the boundary layer thickness varies almost symmetrically. However, along the plane containing LSC, $\delta_{\theta}^{sl}$ shows an asymmetric trend, as it is thicker at one side and thinner near the opposite side of the lateral wall. This is consistent with the time traces of vertical velocity discussed in the previous subsection (refer Fig. \ref{Ra7_Sig2}), wherein, across LSC the signals show high-variance, zero-mean fluctuations and along LSC they exhibit non-zero mean with negative correlation near opposite sides of the lateral wall.

\par
Figure \ref{Bl}(a) shows the comparison between the thermal boundary layer thickness computed by different methods for $Ra=2\times10^7$. The boundary layer thickness remains constant at low rotation rate and increases at higher rotation rates.
We observe that at low rotation rates, boundary layer thickness obtained from all three methods are consistent, while at high rotation rates ($Ro^{-1}\gtrsim10$) the one computed from the rms profile underpredicts the other two.
Similar observation was made by Zhou \emph{et al.} \cite{Zhou_n5_PoF_2011} in two-dimensional RBC, where the thickness from the rms method was found about $20\%$ below that obtained from the slope method.
The variation of azimuthal averaged boundary layer thickness $\langle \delta_{\theta}^{sl} \rangle_{\phi}$ along the radial direction is shown in Figs. \ref{Bl} (b) and (c). 
We find that, the boundary layer is thinner (and almost same as the non-rotating case) at low rotation rates and increases considerably at higher rotation rates.
The effect of rotation is evident near the core region of the cylindrical domain.
However, it is  less apparent near the lateral wall due to the near wall viscous effects. Further, we observe that for low rotation rates, particularly in the LSC regime [red lines in  Figs. \ref{Bl}(b) and (c)], the boundary layer is thicker near the wall and thinner near the core region, while at high rotation rates the trend is reversed.
As rotation rate increases, the flow stabilizes as per Taylor Proudman theorem \cite{taylorp,proudmant} and the temperature field approaches a conduction-like profile.
Here we identify that flow stabilization is reflected noticeably near the core region as $\langle \delta_{\theta}^{sl} \rangle_{\phi}$ is much thicker along the core compared to that near the lateral wall. 

\par
Two quantities which play an important role in local and global heat transport process are the thermal and viscous dissipation rates. 
The global average of dissipation rates are connected to the global heat transport through the analytical relations $\langle\bigepsilon^{th}_u\rangle = \nu^{3}H^{-4} (Nu-1)RaPr^{-2}$ and $\langle \bigepsilon^{th}_{\theta}\rangle=\alpha (\Delta T)^{2}NuH^{-2}$ \cite{Shraiman_Siggia_1990}. 
Dissipation rates are often used to identify different flow regimes \cite{Grossmann_Lohse_JFM_2000, Grossmann_Lohse_2004}, establish scaling laws and quantify different flow characteristics. For instance, Shishkina and Wagner \cite{Shishkina_Wagner_2006,Shishkina_Wagner_2008} investigated the formation and interactions of thermal plumes by evaluating the dissipation rates.

\par
\begin{figure}
\centering
\includegraphics[scale=0.370]{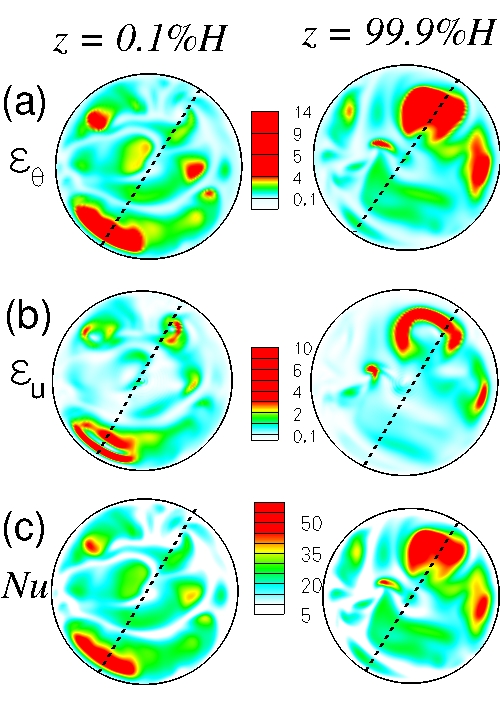}
\caption{(a) and (b) Contours of thermal and viscous dissipation near the bottom (left panel) and top (right panel) plate. The maximum dissipation is observed in the region where the plumes splash into boundary layers near the isothermal plates. (d) Contours of Nusselt number at the bottom (left panel) and top (right panel) plate. The dotted line indicates the azimuthal orientation of LSC.}
\label{LSC_Diss}
\end{figure}

Here we analyze the dissipation rates and their association with the orientation of LSC.
The contours of thermal and viscous dissipation rates near the top and bottom plates [at the same instant as in Fig. \ref{BLT_plot_LSC}(a)] are shown in Figs. \ref{LSC_Diss}(a) and (b), where the azimuthal orientation of LSC is represented by the dashed line. We observe that the dissipation rates vary significantly along the direction of LSC. The thermal and viscous dissipations are maximum in the regions where the plumes collide with the boundary layer. Similar high-amplitude dissipation events associated with the collision of thermal plumes were observed by Schumacher and Scheel \cite{Schumacher_Scheel_2016}. Further, the contours of Nusselt number at the top and bottom plates are shown in  Fig. \ref{LSC_Diss}(c), which also suggests maximum heat transfer occurs at locations where the plume collide with the boundary layer.


\par
\begin{figure}
\centering
\includegraphics[scale=0.3]{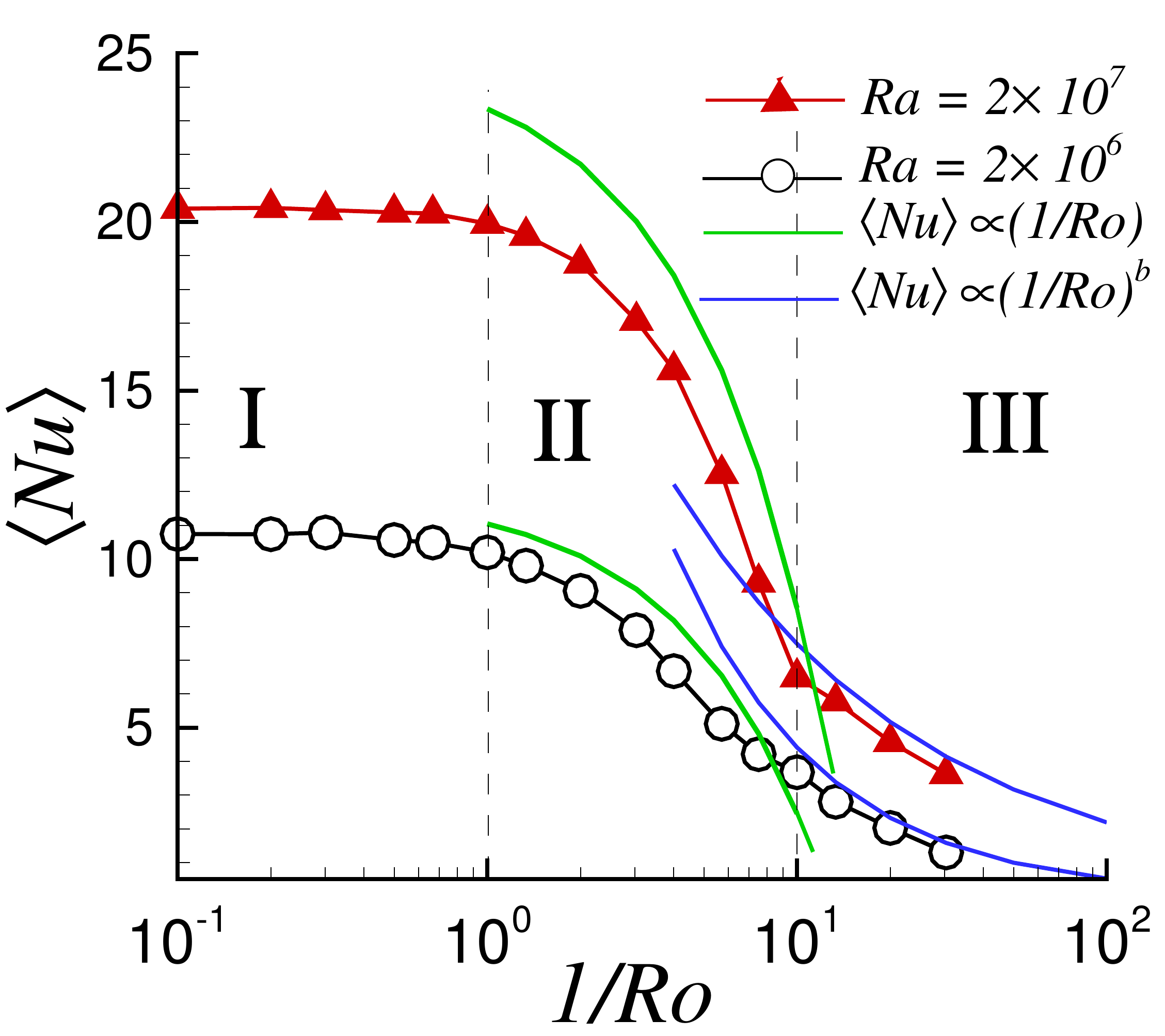}
\caption{(Colour online) Variation of average Nusselt number with rotation rate at $Ra=2\times10^6$ and $2\times10^7$. The solid red and blue lines indicate linear and power law fits, respectively}
\label{Nu_scaling}
\end{figure}

Next we discuss the behaviour of the Nusselt number which is a measure of heat transport, and dissipation rates in different flow regimes. 
Figure \ref{Nu_scaling} shows the variation of average Nusselt number with rotation rate at different Rayleigh numbers.
In regime \Romannum{1} ($0 \leq Ro^{-1} \lesssim 1$) or LSC regime, heat transfer remains almost constant. As rotation rate increases Nusselt number drops linearly as $\langle Nu \rangle \propto  1/Ro$ in regime \Romannum{2} given by $1 \lesssim Ro^{-1} \lesssim 10$. However, at high rotation rates ($Ro^{-1} \gtrsim 10$, regime \Romannum{3}) the Nusselt number shows a power law behaviour, $\langle Nu \rangle \propto (1/Ro)^{b}$, with the exponent $b=-0.93$ and $-0.53$ for $Ra=2\times10^6$ and $2\times10^7$, respectively.  The solid red and blue lines in Fig. \ref{Nu_scaling} represent the corresponding linear and power law fits at regimes \Romannum{2} and \Romannum{3}, respectively.
%

\par
\begin{figure}
\centering
\includegraphics[scale=0.3]{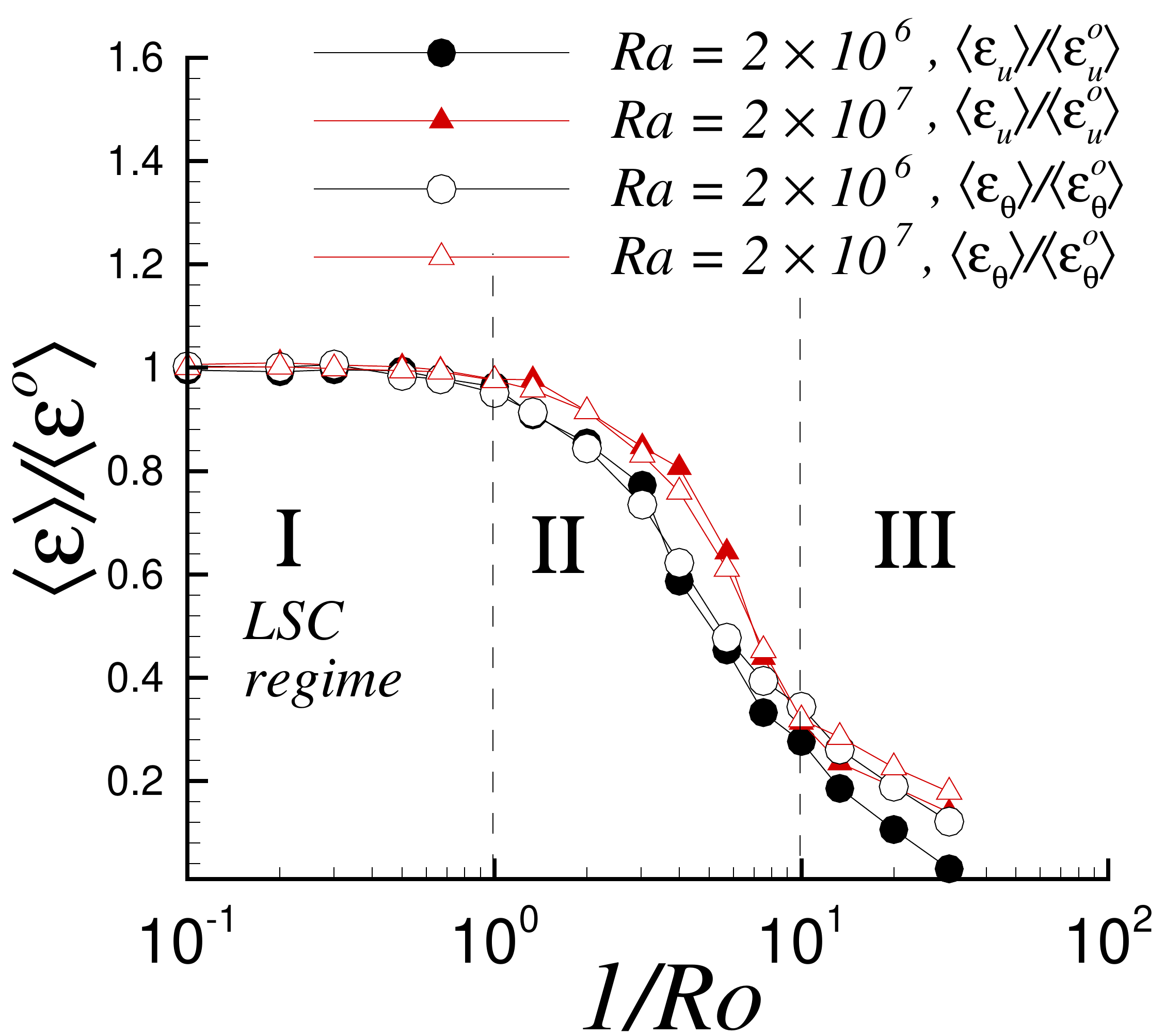}
\caption{(Colour online) Variation of $ \langle\bigepsilon_{u}\rangle / \langle\bigepsilon_u^{o}\rangle$  and $\langle\bigepsilon_{\theta}\rangle /\langle\bigepsilon_{\theta}^{o}\rangle$ with rotation rate at $Ra=2\times10^6$ and $2\times10^7$.}
\label{Diss_scale}
\end{figure}

Figure \ref{Diss_scale} shows the ensemble averaged non-dimensional dissipation rates, $\langle\bigepsilon_u\rangle =\sqrt{Pr/Ra} \langle\lvert\nabla \mathbf{u} \rvert ^{2}\rangle$ and $\langle\bigepsilon_{\theta}\rangle=\sqrt{1/RaPr} \langle\lvert\nabla \theta \rvert^{2}\rangle$, normalized by the corresponding non-rotating dissipation rates $\langle\bigepsilon_u^{o}\rangle$ and $\langle\bigepsilon_{\theta}^{o}\rangle$, respectively, for different rotation rates. In regime \Romannum{1} where LSC is observed both dissipation rates remain almost constant and close to the non-rotating value.
However, at higher rotation rates a significant departure from the non-rotating case is noticed. 
At regime \Romannum{2} the dissipation rates decrease linearly with rotation rate, as observed for the Nusselt number. This is followed by rotation dominated regime \Romannum{3} where they follow a power law behaviour as  $\langle\bigepsilon\rangle \propto (1/Ro)^{b}$, with exponent $b=-1.95$ and $-0.68$ for $Ra=2\times10^6$ and $2\times10^7$, respectively, for  $\langle\bigepsilon_u\rangle$, while $\langle\bigepsilon_{\theta}\rangle$  follows the same power law as that of  $\langle Nu \rangle$.
\par
\begin{figure}
\centering
\includegraphics[scale=0.330]{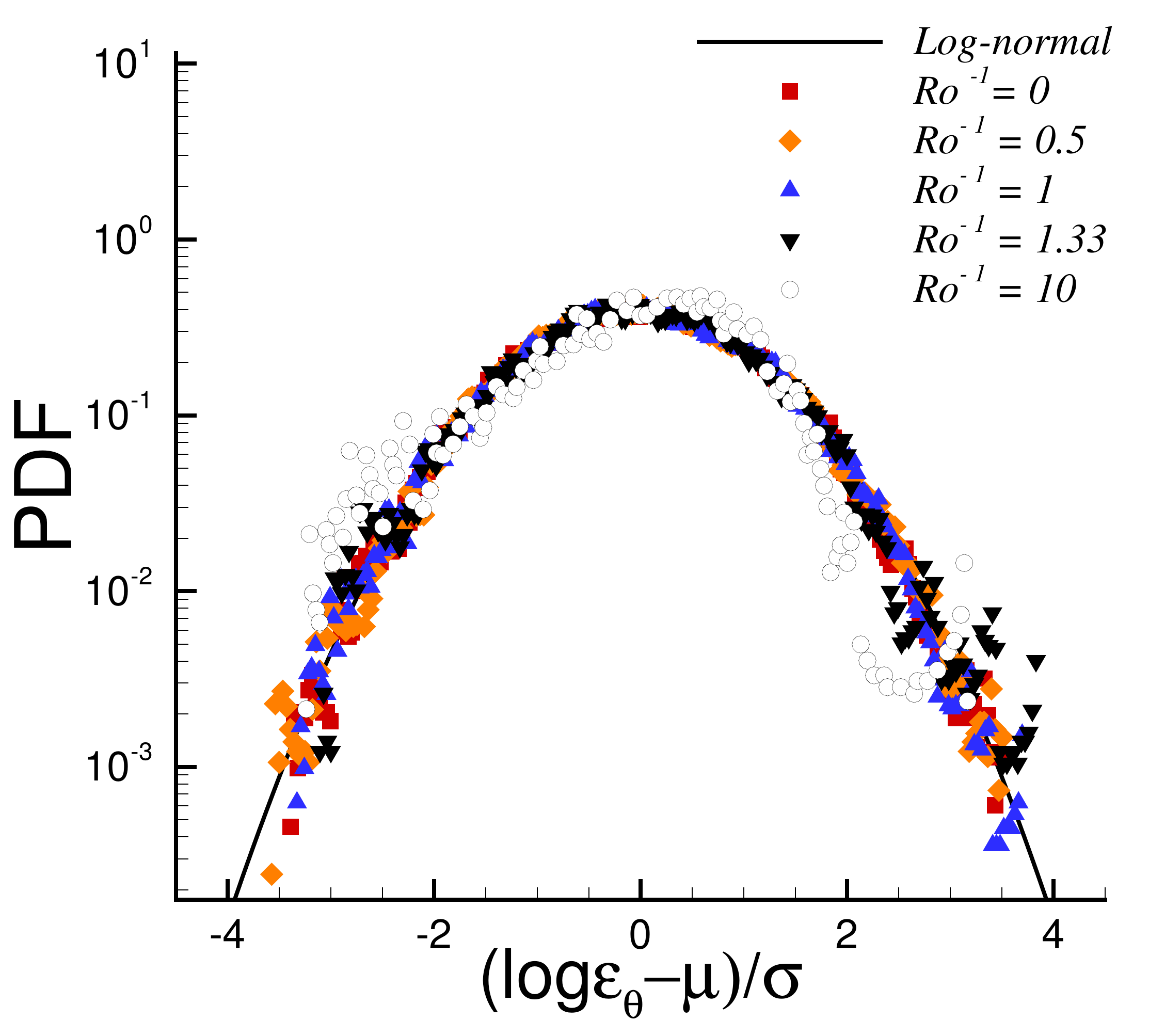}
\caption{(Colour online) PDF of log$\epsilon_{\theta}$ at  $Ra=2\times10^7$ for different $Ro^{-1}$ in log-normal coordinates.}
\label{DissEt_PDF}
\end{figure}

In RBC, dissipation fields are often expected to follow a log-normal distribution \cite{Kolmogorov_1962}. However, considerable deviations due to the highly intermittent nature of the local dissipation has also been reported \cite{Zhang_Zhou_SunJFM2017,Emran_Schumacher_2008,Schumacher_Sreenivasan_2005,Kaczorowski_Wagner_2009}. In Fig. \ref{DissEt_PDF}, PDF of thermal dissipation rate is represented in log-normal coordinates, where $\mu$ and $\sigma$ indicate the mean and standard deviation of log $\bigepsilon_{\theta}$. 
In the LSC regime, both $\bigepsilon_{u}$ and $\bigepsilon_{\theta}$ follow the log-normal behaviour. However at regime \Romannum{3}, a clear departure from log-normality is noticed ($Ro^{-1} = 10$).
Further, the tails of PDF become shorter at regime \Romannum{3}, implying a reduced probability of high-amplitude dissipation events.


\begin{figure}
\centering
\includegraphics[scale=0.375]{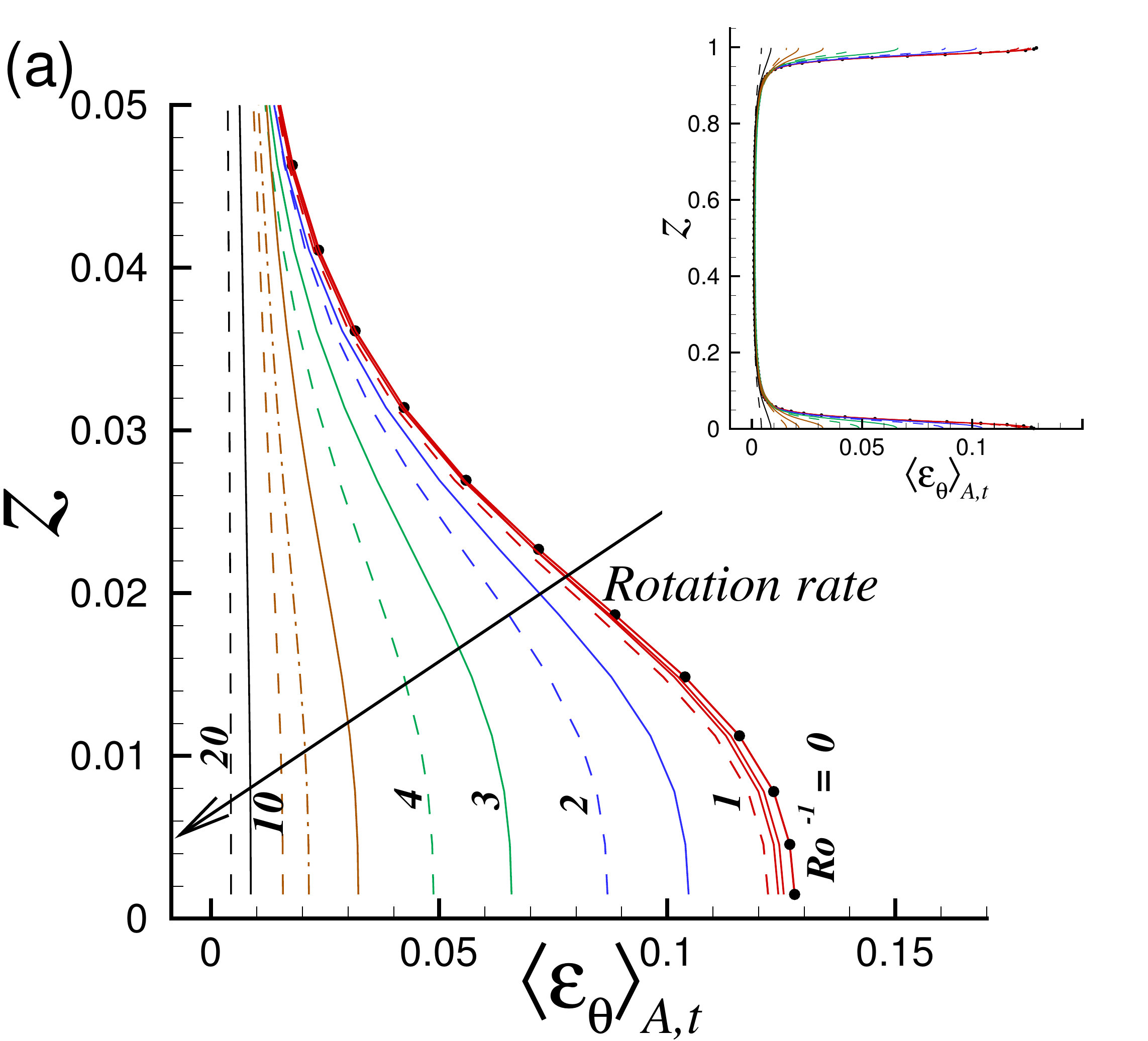}\\
\includegraphics[scale=0.375]{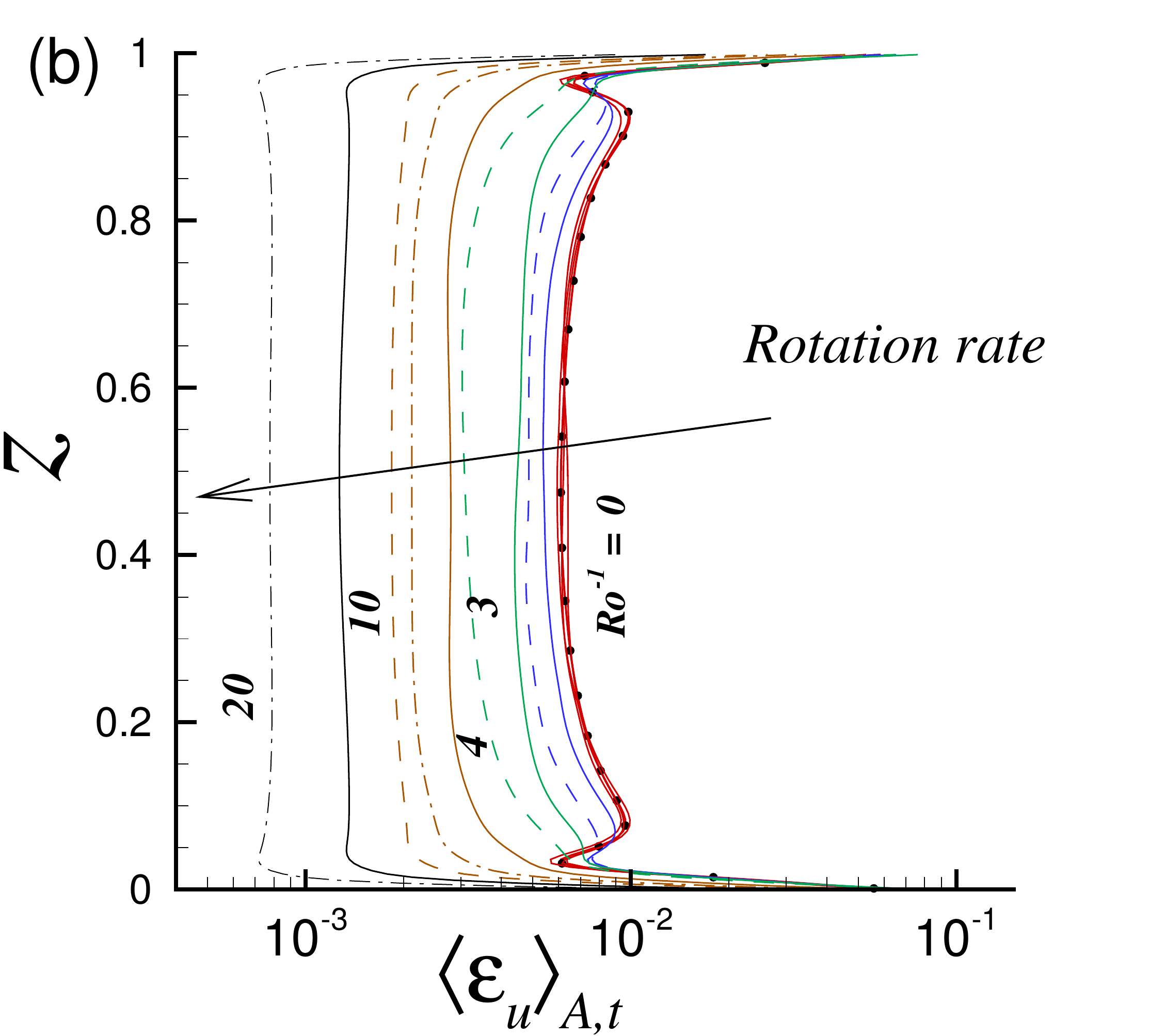}
\caption{(Colour online) (a) Enlarged view (near bottom plate) of the variation of thermal dissipation along the vertical direction at $Ra=2\times10^6$ for different $Ro^{-1}$. The inset shows the complete profile. (b) Variation of average viscous dissipation rate along the vertical direction for different rotation rates in semi-log scale. The inset shows the zoomed view of the same near the bottom plate. Arrows indicate the direction of increase in rotation rate.}
\label{Diss_Axial}
\end{figure}

\par
We now examine the spatial distribution of the viscous and thermal dissipation rates in turbulent RRBC.
The variation of the horizontal plane and time averaged dissipation rates along the vertical direction are shown in Fig. \ref{Diss_Axial}, where the complete profile of  $\langle \bigepsilon_{\theta} \rangle_{A,t}$ is shown in the inset (a). 
We observe that maximum value of the dissipation rates occur at the top and bottom plates which is consistent with the observations made by  Zhang \emph{et al.} \cite{Zhang_Zhou_SunJFM2017} and Emran and Schumacher \cite{Emran_Schumacher_2012}. 
In the LSC regime, the bulk region shows negligible thermal dissipation (in comparison to boundary layers), indicating most of the thermal energy is dissipated near the thermal boundary layers. As rotation rate increases (regime \Romannum{3}) the magnitude of thermal dissipation decreases near the top and bottom surfaces. The straight vertical lines in regime \Romannum{3} suggest that the energy is dissipated more uniformly throughout the domain.
On a similar note, the viscous dissipation also follows a different profile in the LSC regime compared to that at high rotation rates. In the LSC regime two local peaks are identified in the $\langle \bigepsilon_{u} \rangle_{A,t}$ profile before it becomes stable near the bulk region. However at regime \Romannum{3}, only a single peak is identified. As rotation rate increases the average dissipation rates decrease and, $\langle \bigepsilon_u \rangle_{A,t}$ and $\langle \bigepsilon_{\theta} \rangle_{A,t}$ approaches steady profiles, as all the velocity and temperature fluctuations in the system are damped out. The stabilizing effect of rotation on RBC is evident from these profiles.  In the following subsection we discuss the dynamics of LSC, mainly the changes in the azimuthal direction and further present their statistics. 


\subsection{Dynamics of reorientations}
The vertical plane containing LSC is known  
to show spontaneous and erratic drifts or directional changes with time \cite{Keller_1966,Welander_1967,Creveling_etal_1975,Gorman_etal_1984,Hansen_etal_1992}.
These sudden and significant changes in the orientation of LSC are referred as reorientations of LSC \cite{Brown_Ahlers_JFM}. 
During some reorientations the amplitude of the first Fourier mode almost drops to zero and such events are termed as cessation-led reorientations. While in others, the LSC rotates azimuthally  without considerable change in the amplitude of the first Fourier mode \cite{Cioni_etal_1997,Brown_Nikolaenko_Ahlers_PRL_2005}. These are referred as rotation-led reorientations. 
Although these events are reported experimentally by Brown \emph{et al.} \cite{Brown_Nikolaenko_Ahlers_PRL_2005}, Brown and Ahlers \cite{Brown_Ahlers_JFM} and  Xi \emph{et al.} \cite{Xi_Zhou_Xia_2006}, confirmation by numerical studies is rare. 
Here we probe the dynamics and statistics of reorientations of LSC in RRBC using direct numerical simulations. 

\par
Reorientations are generally quantified using the amplitude $A_{k}=|\hat{u}(k)|$ and phase $\Phi_{k}= tan^{-1} (\mathbf{Im~} \hat{u}(k) / \mathbf{Re~}\hat{u}(k))$ of the Fourier modes (refer to  Eqn. \ref{gov_eqn4}). The characterization of reorientations of LSC is not a well established concept, but in literature we find few works in this direction. For example, Brown \emph{et al.} \cite{Brown_Nikolaenko_Ahlers_PRL_2005} and Brown and Ahlers \cite{Brown_Ahlers_JFM} identified the reorientations based on the phase of the first Fourier mode $\Phi_1$ using two criteria:
(a) The net angular change $|\Delta \Phi_1|$ has to be greater than $\pi/4$, and (b) magnitude of the net azimuthal rotation rate $|\Delta \Phi_1/\Delta t|$ has to be greater than $\pi/5T_{eddy}$, where $T_{eddy}$ is the eddy turn over time calculated as $T_{eddy}=2H/w^{rms}$. Here $w^{rms}$ is the average value of the rms of vertical velocity time signals from the previously mentioned $36$ numerical probes placed at the mid-vertical plane. 
The above procedure results in multiple overlapping subsets of reorientations and thus requires further conditioning using quality factors to pick the appropriate reorientation. To avoid this complexity, in the present study we ensure that the azimuthal rotation rate between any two adjacent points is greater than the specified cutoff ($\pi/5T_{eddy}$). 
%
Further, we identify these reorientations as
complete reversal if the phase change $|\Delta \Phi_1| = \pi$ and as partial reversal if $|\Delta \Phi_1|\neq \pi$, as proposed by Mishra \emph{et al.} \cite{MishraDe}.

\begin{figure}
\centering
\includegraphics[scale=0.38]{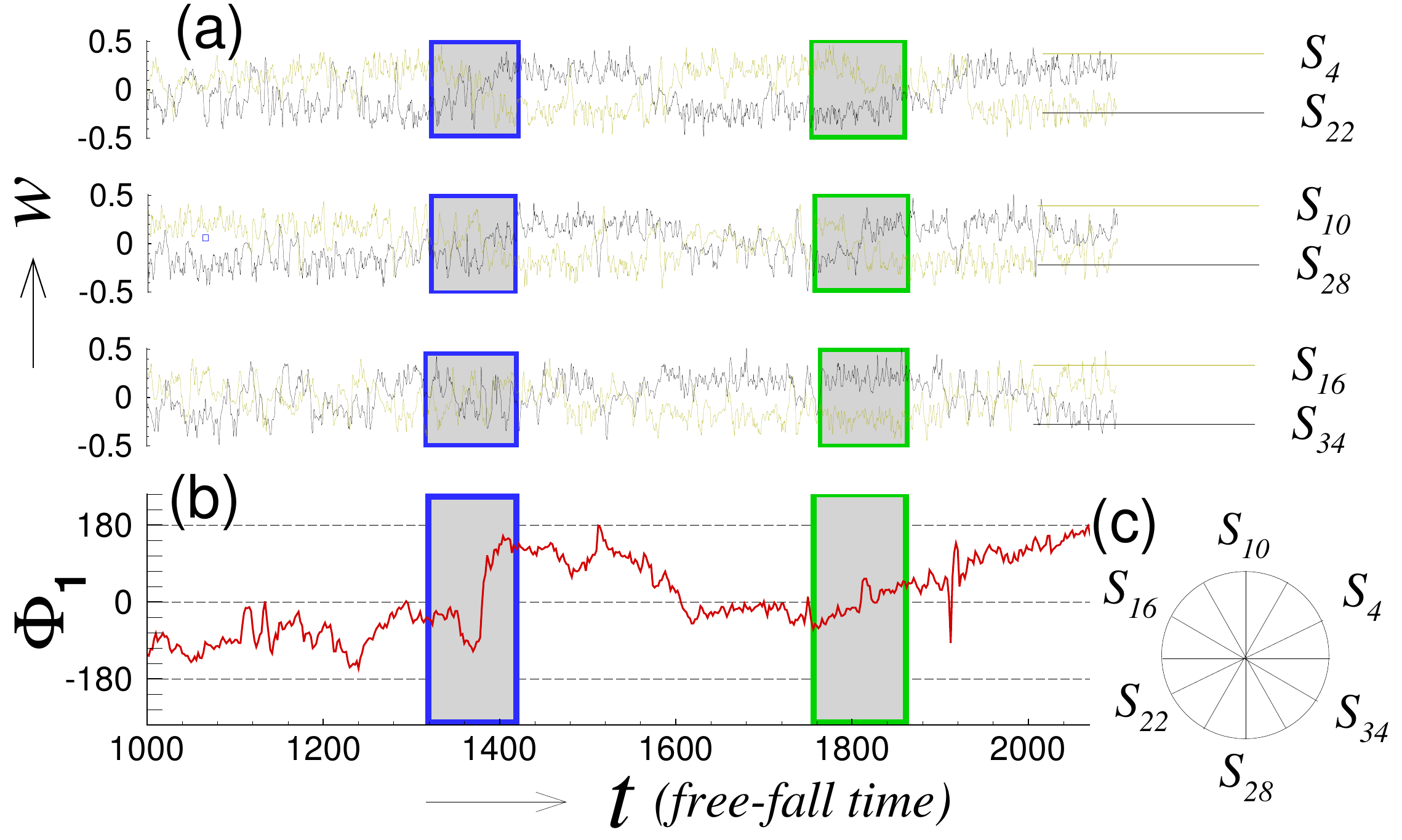}
\caption{(Colour online) (a) Time signals of vertical velocity from different numerical probes at mid-plane of the cylinder and (b) the phase of first Fourier mode showing complete (blue box) and partial reversals (green box) for $Ra=2\times10^6$ and $Ro^{-1}=0$. (c) Schematic representation of the location of the numerical probes.}
\label{Rot_Re_sig}
\end{figure}


\par
The time evolution of the phase of the first Fourier mode along with the vertical velocity signals from different numerical probes are shown in Fig. \ref{Rot_Re_sig}.
Here we identify reversals by the sudden and considerable change in the phase $\Phi_1$, along with the switching of mean vertical velocity.
At $t\approx 1380$, a complete reversal is observed, where the phase change $|\Delta \Phi_1| \approx \pi$. This is accompanied by the switching of mean flow across all the probes. However, for the reversal at $t\approx 1810$, the mean value of $w$ does not switch at all probes. Clearly, at probes $S_{4}~(\phi=\pi/6),~ S_{22}~(\phi=7\pi/6),~S_{16}~(\phi=5\pi/6)$ and $S_{34}~(\phi=11\pi/6)$ the mean vertical velocity remains unaltered, while it switches sign at $S_{10}~(\phi=\pi/2)$ and $S_{28}~(\phi=3\pi/2)$. This is a partial reversal where the phase change is less than $\pi$. We observe several partial reversals in our simulations while the number of complete reversals are very less.


\begin{figure}[h!]
\centering
\includegraphics[scale=0.350]{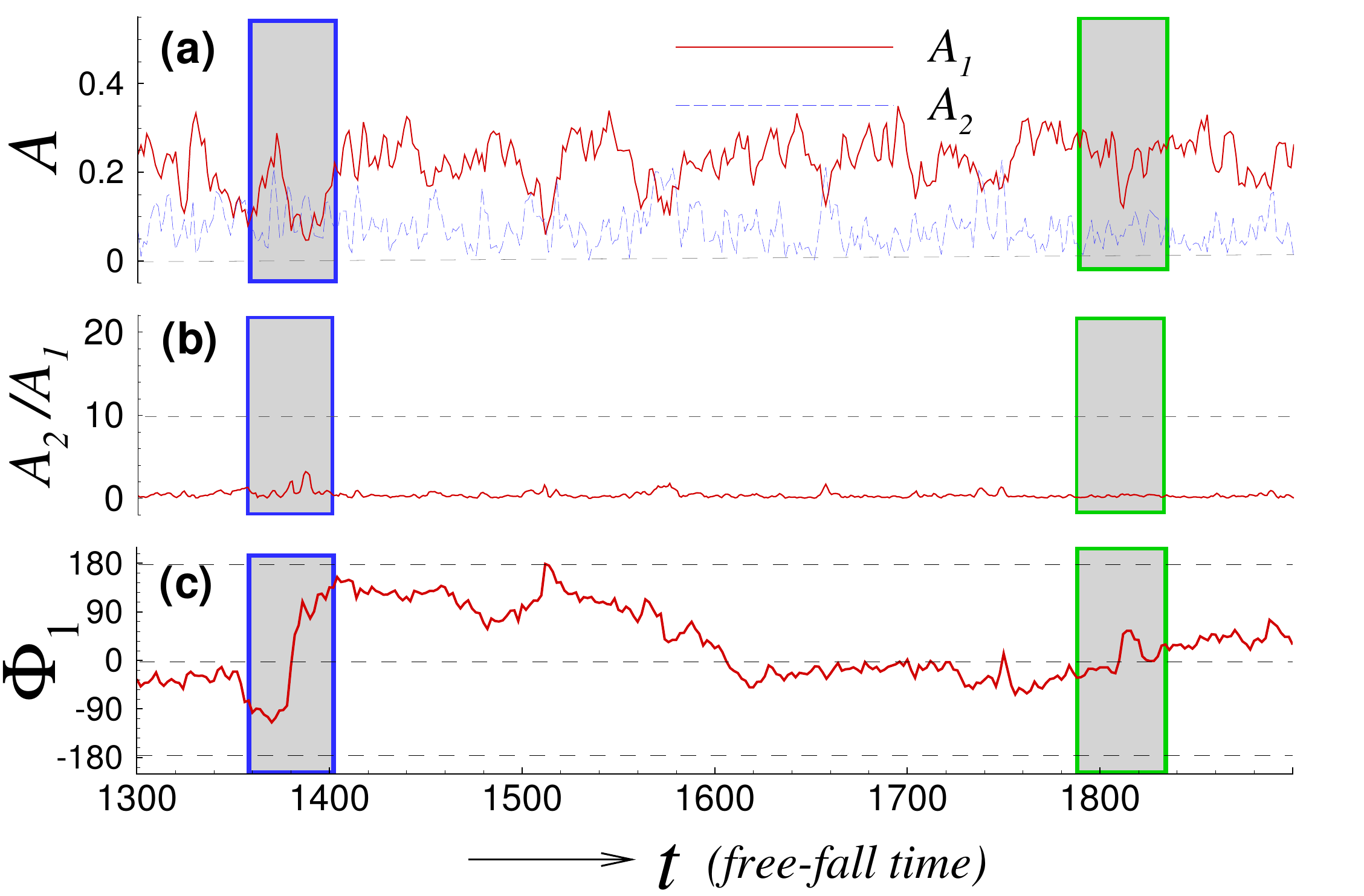}
\caption{(Colour online) The time series of first and second Fourier modes, their corresponding amplitude ratio $A_{2}/A_{1}$ and phase of the first Fourier mode during complete (blue box) and partial (green box) reversals at  $Ra=2\times10^6$ and $Ro^{-1}=0$.}
\label{CP_reveRa6i}
\end{figure}

\par
To get better insights into the nature of these reversals we analyze the  time evolution of the first and second Fourier modes. We observe that for RBC and RRBC at low rotation rates (where LSC persists) the first Fourier mode dominates over the other modes. In Fig. \ref{CP_reveRa6i}, we observe that during the reorientations amplitude of the first Fourier mode $A_1=|\hat{u}_{1}|$ drops below its mean value. On closer inspection, although $A_1$ shows a small dip in magnitude near $t\approx1380$, the rise in the amplitude of the second mode $A_2=|\hat{u}_{2}|$ is marginal and thus the amplitude fraction $A_{2}/A_{1}$ remains negligibly small. Thus we identify these as rotation-led reorientation, where the LSC structure rotates azimuthally to reorient itself to a new direction without much change in the circulation strength. 

\par
\begin{figure}
\centering
\includegraphics[scale=0.40]{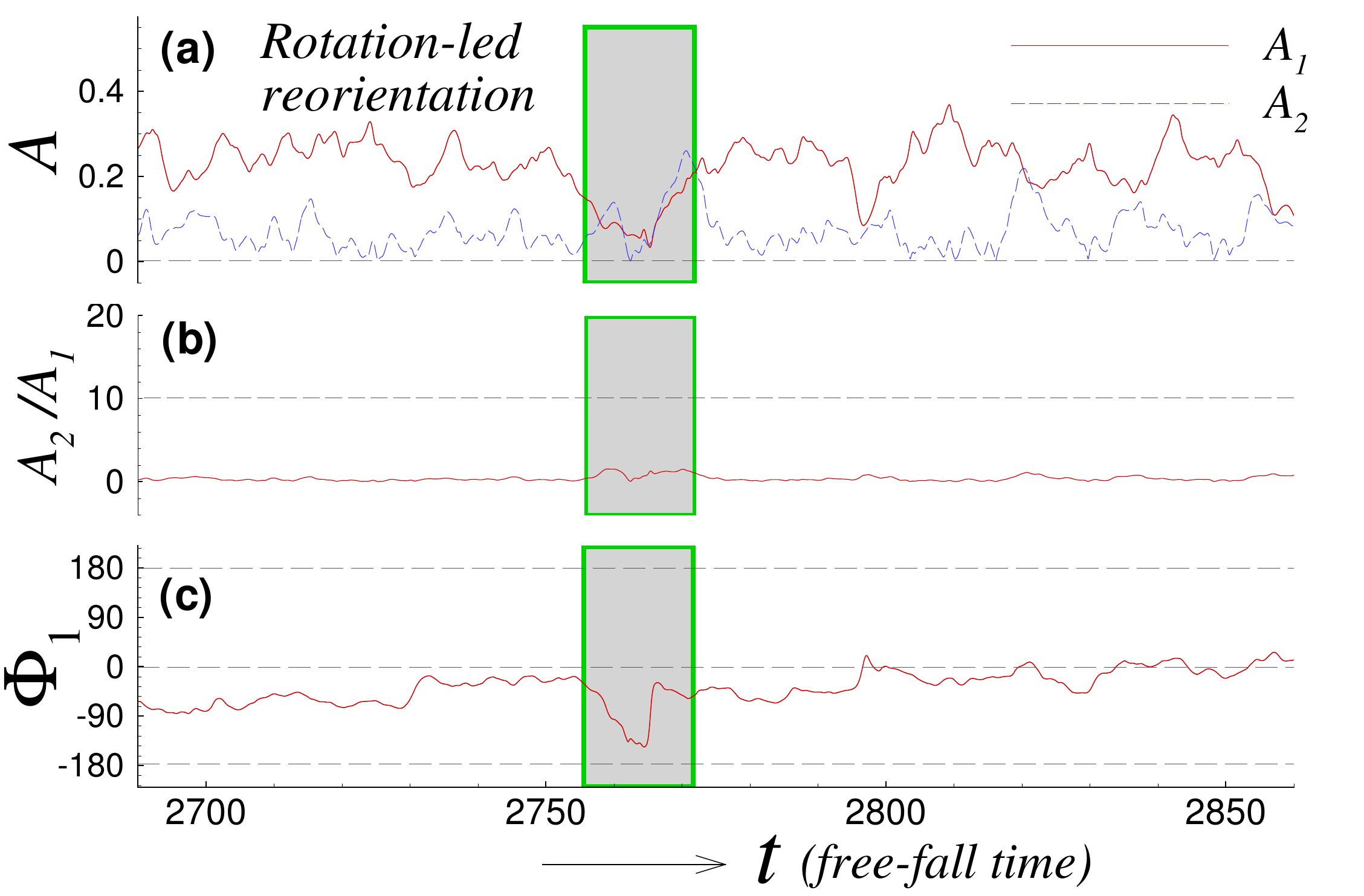}
\caption{The time series of first and second Fourier modes, their corresponding amplitude ratio $A_{2}/A_{1}$ and phase of the first Fourier mode during rotation-led partial reversal obtained for $Ra=2\times10^6$ and $Ro^{-1}=0.1$. The boxed region identifies the reorientation. }
\label{P_Rot}
\end{figure}
\begin{figure}
\centering
\includegraphics[scale=0.40]{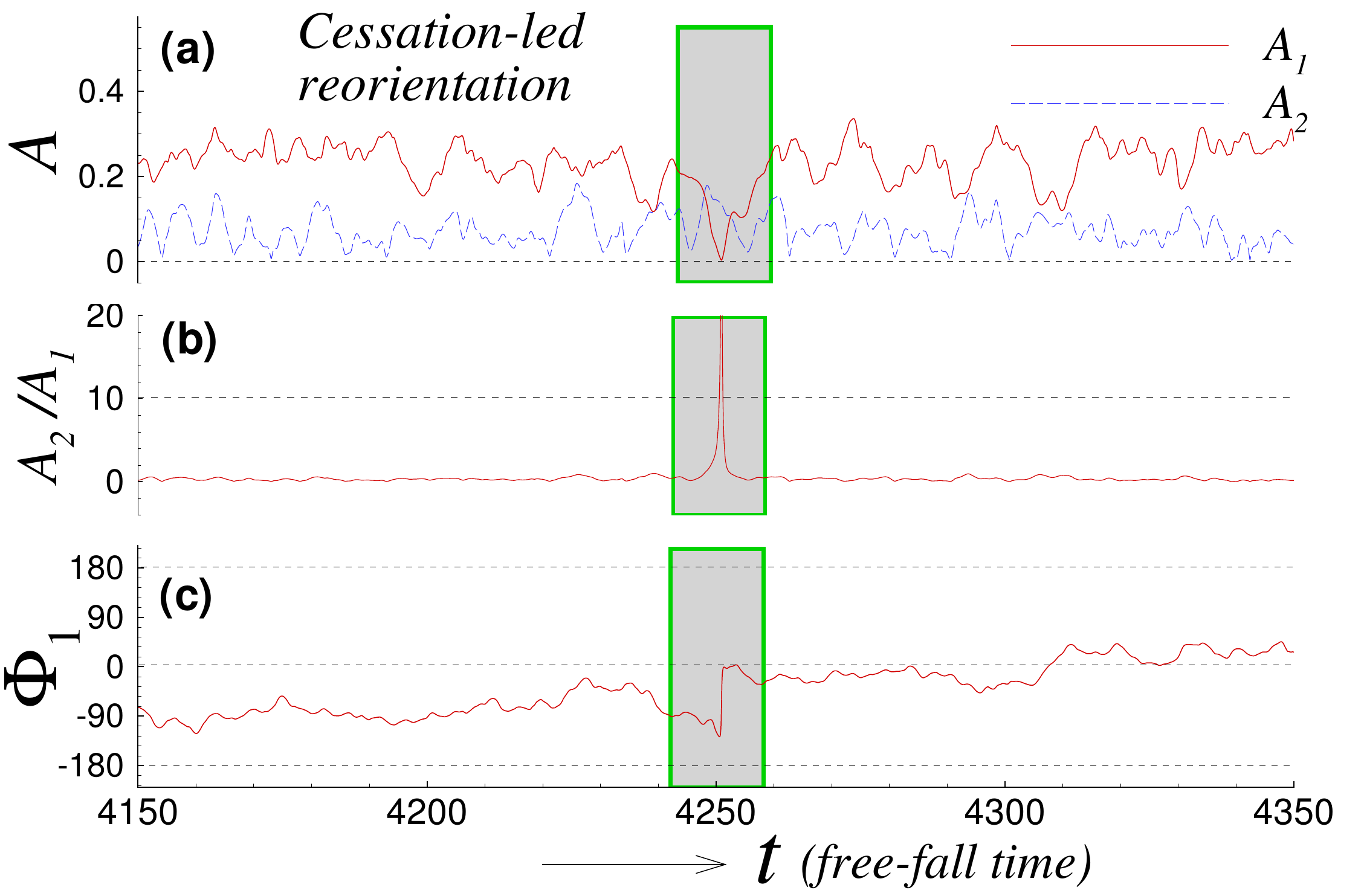}
\caption{The time series of first and second Fourier modes, their corresponding amplitude ratio $A_{2}/A_{1}$ and phase of the first Fourier mode during cessation-led reorientation obtained for $Ra=2\times10^6$ and $Ro^{-1}=0.2$. The dotted line in the amplitude fraction plot indicates the cutoff value $A_{2}/A_{1}=10$, used to identify cessations. }
\label{C_Rot}
\end{figure}

Rotation-led reorientation observed for  rotating RBC is shown in Fig. \ref{P_Rot} for $Ra=2\times10^6$ and $Ro^{-1}=0.1$.
 At $t\approx 2765$ the phase of the first Fourier mode shows an angular change $|\Delta\Phi_1|\approx 2\pi/3$. Although, $A_1$ shows a local minimum at this instant, $A_2$ is also minimal and resultantly the amplitude fraction is close to zero as shown in frame (b). 
However, during a different reorientation (refer Fig. \ref{C_Rot}) the amplitude of the first Fourier mode drops considerably ($A_{1} \rightarrow 0$) and that of the second rises. These are identified as cessation-led reorientations. Here the strength of LSC  diminishes at the cessation point (peak of $A_{2}/A_{1}$) and the flow reorients itself at any arbitrarily chosen direction. Due to the fluctuations in the $A_1$ and $A_2$ time signals, we use the amplitude fraction $A_{2}/A_{1}$ to clearly recognize the cessations. In the present study cessations are identified as sharp peaks in the amplitude fraction such that $A_{2}/A_{1}\gtrsim10$. 
Here we observe a a partial cessation-led reorientation near $t\approx 4250$ with an angular change $|\Delta \Phi_1| \approx 3\pi/4$. 
The experimental studies by  Brown \emph{et al.} \cite{Brown_Nikolaenko_Ahlers_PRL_2005}, Brown and Ahlers \cite{Brown_Ahlers_JFM} and Xi \emph{et al.} \cite{Xi_Zhou_Xia_2006}  have reported similar reorientations. 

\par
\begin{figure}[t!]
\centering
\includegraphics[scale=0.39]{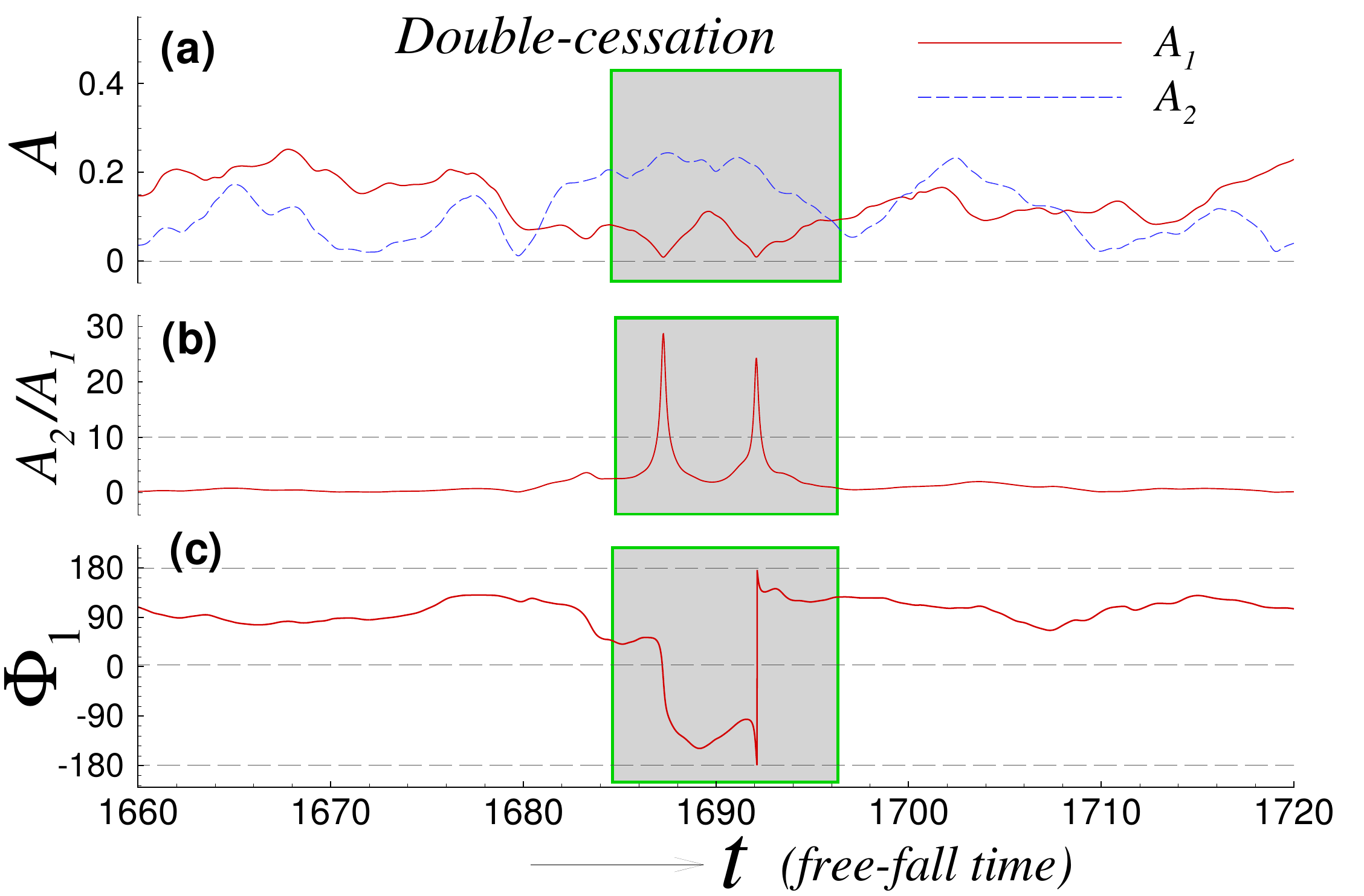}\\
\caption{The time series of first and second Fourier modes, their corresponding amplitude ratio $A_{2}/A_{1}$ and phase of the first Fourier mode during double  cessation observed for $Ra=2\times10^6$ at $Ro^{-1}= 0.3$. The dotted line in the amplitude fraction plot indicates the cutoff value $A_{2}/A_{1}=10$, used to identify cessations. The boxed region bounds an eddy turnover time within which the double-cessation prevails. }
\label{Cess_D}
\end{figure}

Occasionally, two cessations occur in quick succession, i.e., within an eddy turnover time. These are referred as double-cessations.
Figure \ref{Cess_D} shows a double-cessation observed for $Ra=2\times10^6$ and $Ro^{-1}=0.3$. The boxed region bounds an eddy turnover time ($T_{eddy}\approx 11$ free-fall time units), within which the double-cessation occurs. The amplitude of the first Fourier mode drops ($A_{1} \rightarrow 0$) at $t\approx1687$ and $t\approx1692$, accompanied by sharp rise in the amplitude fraction $A_{2}/A_{1}$. Here we observe that the phase of first Fourier mode $\Phi_{1}$ changes successively, first by $|\Delta \Phi_1| \approx \pi$ and second by $|\Delta \Phi_1 |\approx \pi/2$. The net angular change as a result of the double-cessation is accounted as the sum of these two successive events. Similar double-cessations were observed experimentally by Xi \emph{et al.} \cite{Xi_Zhou_Xia_2006}, and were later numerically reported by Mishra \emph{et al.} \cite{MishraDe}. The number of double-cessations observed is much fewer compared to regular (or single) cessations.
\par
\begin{figure}[t!]
\centering
\includegraphics[scale=0.39]{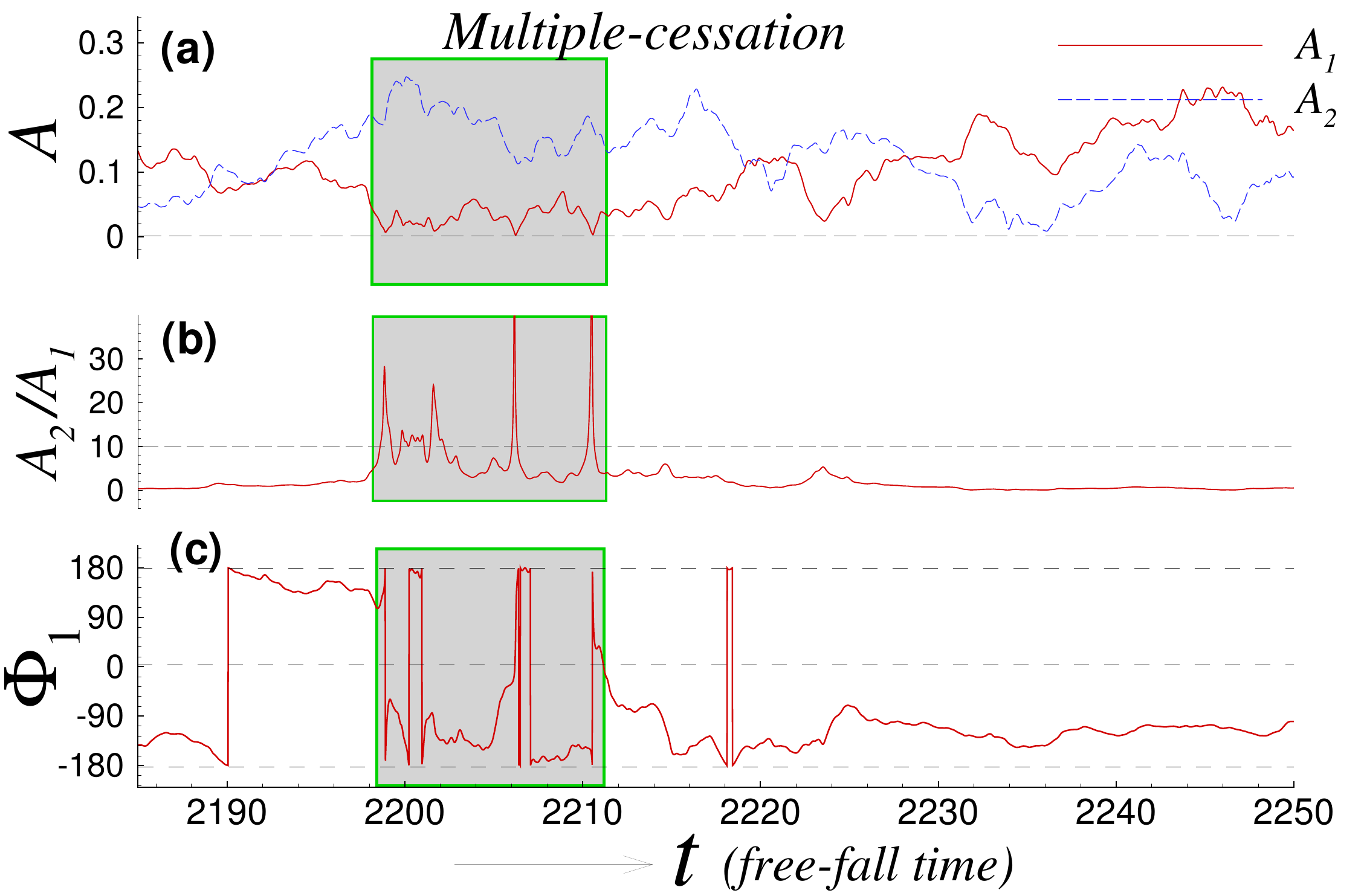}
\caption{The time series of first and second Fourier modes, their corresponding amplitude ratio $A_{2}/A_{1}$ and phase of the first Fourier mode during multiple cessations observed for  $Ra=2\times10^7$ at $Ro^{-1}=0.2$. The dotted line in the amplitude fraction plot indicates the cutoff value $A_{2}/A_{1}=10$, used to identify cessations. The boxed region bounds an eddy turnover time within which the multiple-cessation prevails.}
\label{Cess_M}
\end{figure}

Interestingly, in our simulations, we observe more than two cessations occur within an eddy turnover time. This has not been reported before and we term this phenomenon as multiple-cessation.  Figure \ref{Cess_M} shows the multiple-cessation observed for $Ra=2\times10^7$ at $Ro^{-1}=0.2$. The boxed region bounds an eddy turnover time within which the multiple-cessation occurs. We notice that $A_1$ drops close to zero successively at $t\approx2199, 2201$, $2205$ and $2210$, and the amplitude fraction $A_{2}/A_{1}$ shows sharp peaks at these locations. The discontinuities (sudden fluctuations) in the phase are jitters near $-\pi$ and $\pi$.  
We observe only a single instance of multiple-cessation in our entire set of simulation.

\par
\begin{figure}
\centering
\includegraphics[scale=0.150]{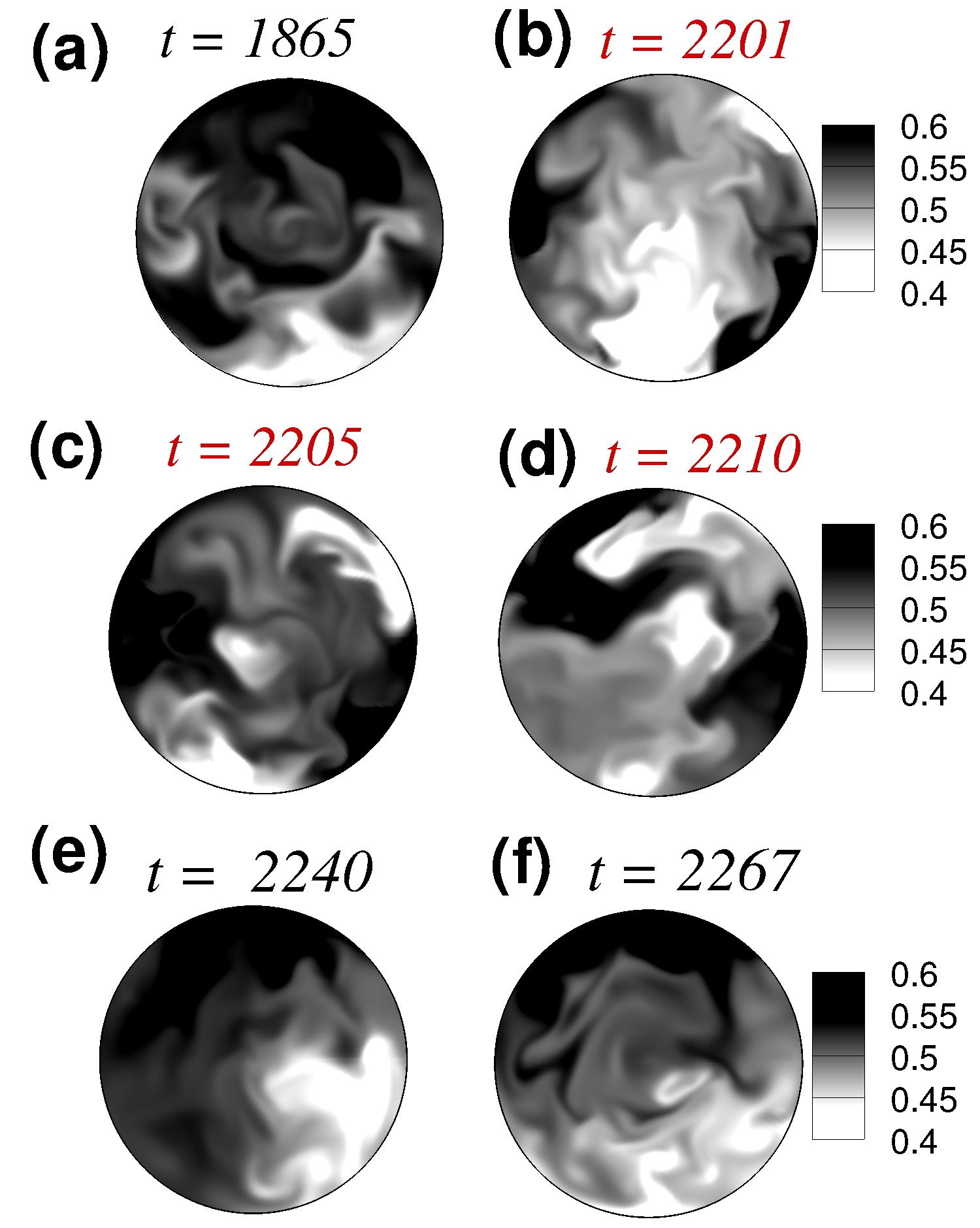}
\caption{Temperature contours at mid-vertical plane showing quadrupolar and dipolar structures observed before ($t=1865$), during ($t=2201,2205$ and $2210$) and after ($t=2240$ and $2267$) cessation, for $Ra=2\times10^7$ at $Ro^{-1}=0.2$. }
\label{Cess_Contours}
\end{figure}

Next we study the relation between cessations and the dynamics of flow structure. We observe that during cessations, as the amplitude of first Fourier mode diminishes the flow structure corresponding to the first mode (dipolar structure) becomes weak, and the quadrupolar structure corresponding to the second Fourier mode become dominant. However, after the cessation, the dipolar structure reappears accompanied by a change in the azimuthal orientation of LSC. This is consistent with the observations made by  Mishra \emph{et al.}\cite{MishraDe} for non-rotating RBC. Here we substantiate this by illustrating the temperature contours at mid-plane ($z=0.5$) before, during and after the cessation, as shown in Fig. \ref{Cess_Contours}. Before the cessation, the contour at $t\approx1865$ follows an almost dipolar structure, where hot and cold fluid appear at opposite sides ($\pi$ apart) of the lateral wall. However, during cessations, i.e., at $t\approx2201$, $2205$ and $2210$ (refer Fig. \ref{Cess_M}  also) the temperature contours indicate a quadrupolar behaviour. Here the hot and cold fluid are observed alternately and separated azimuthally by about $ \pi/2$. After the cessations the flow goes back to the dipolar structure as shown in  Figs. \ref{Cess_Contours} (e) and (f).

 \subsection{Statistics of reorientations}
\par
\begin{figure}[h!]
\centering
\includegraphics[scale=0.19]{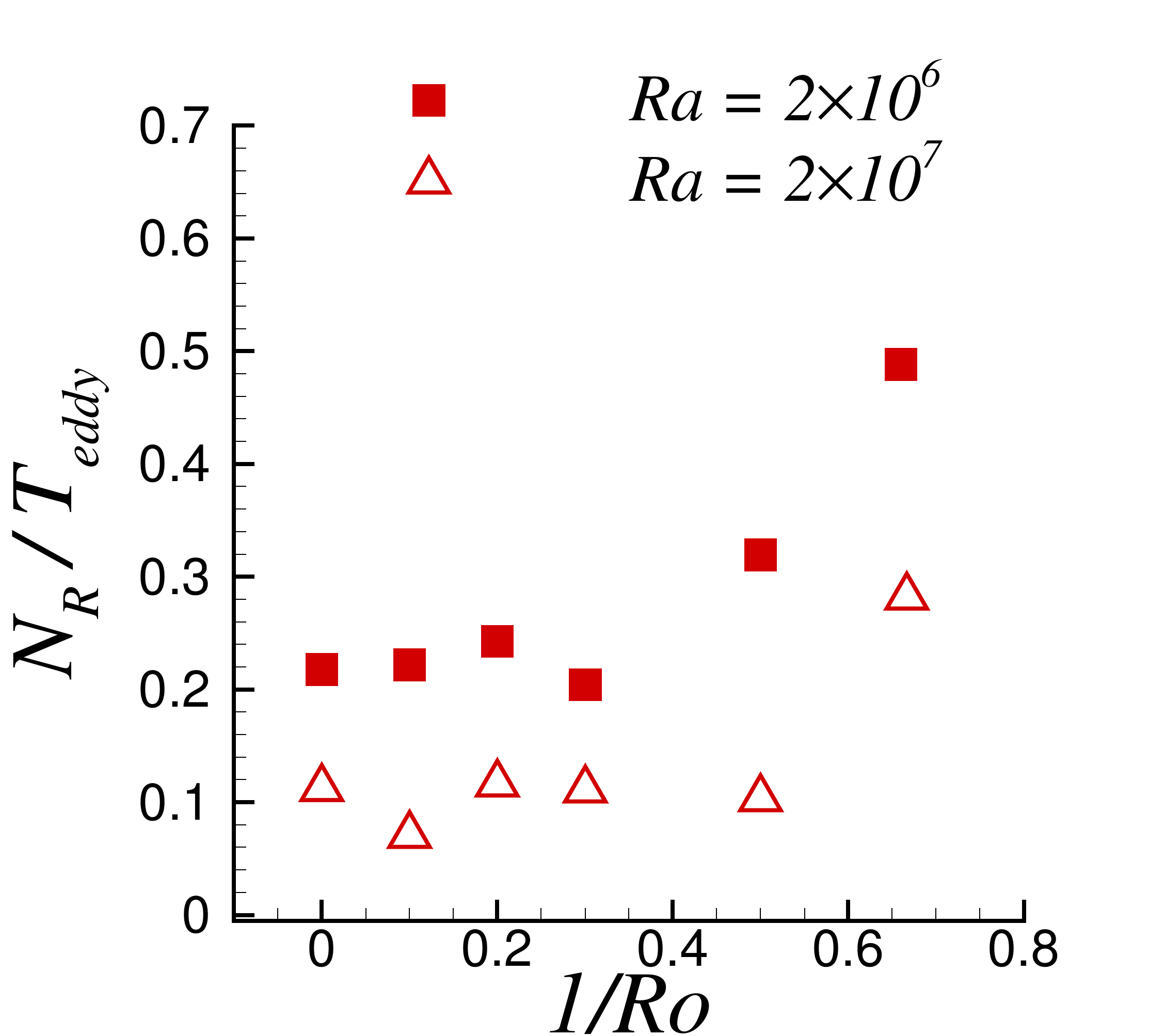}\includegraphics[scale=0.19]{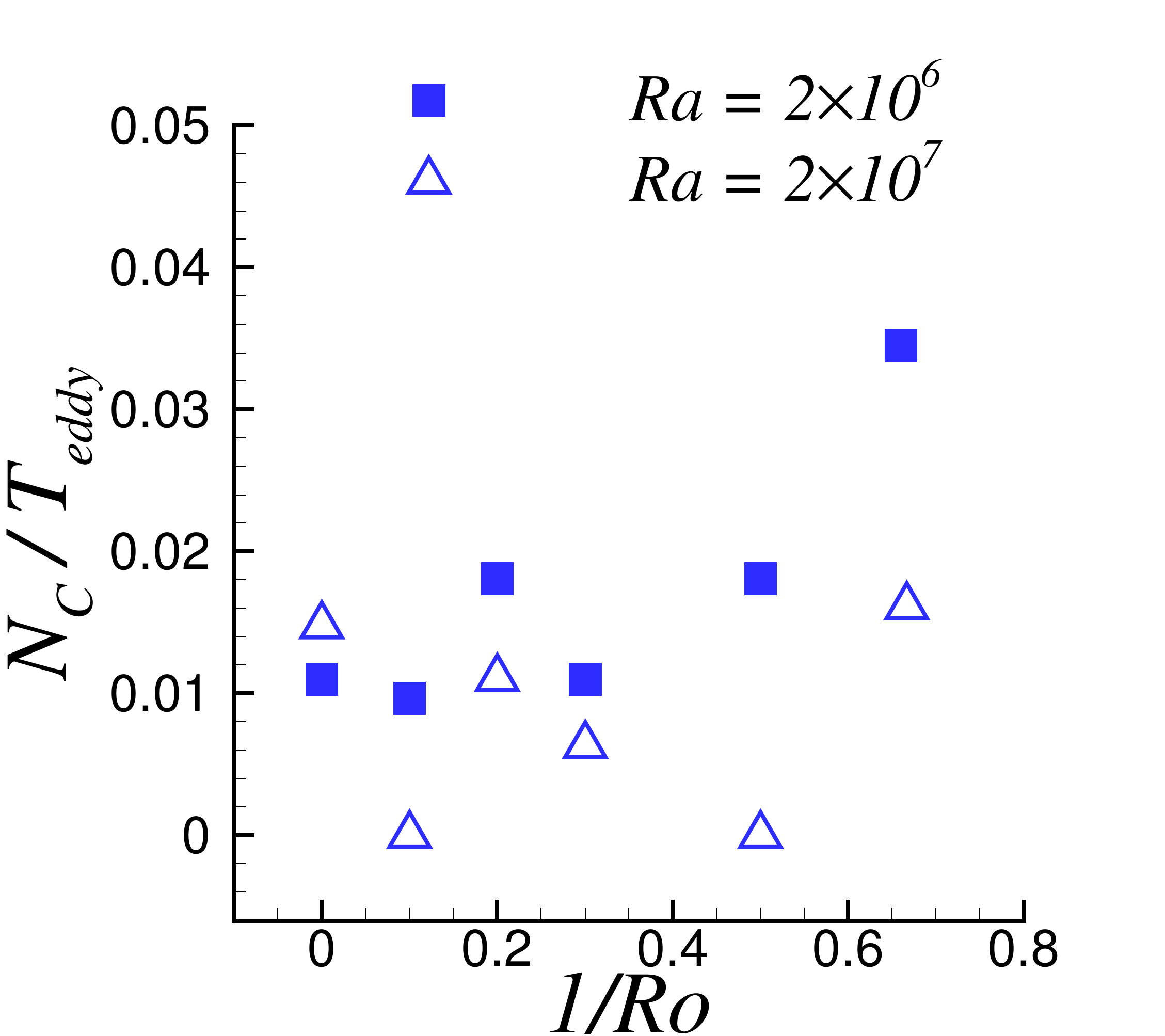}
\caption{Number of reorientations and cessations per eddy turnover time observed for $Ra=2\times10^6$ and $2\times10^7$. }
\label{Reore_stats}
\end{figure}

After investigating the dynamics of reorientations of LSC, we turn our focus on its statistics.
The number of reorientations ($N_R$) and cessations ($N_C$) per eddy turnover time at different rotation rates are shown in Fig. \ref{Reore_stats} for the two Rayleigh numbers.
On an average we consider about $550$ and $300$ eddy turnover times ($T_{eddy}$) for $Ra=2\times10^6$ and $2\times10^7$, respectively. 
With the increase in rotation rate we observe a considerable increase in the number of reorientations.
Further, the number of reorientations are higher for lower $Ra$ for both rotating and non-rotating RBC. Although, we observe comparatively more cessations for lower $Ra$, the number of cessations are quite low to arrive at any conclusion.  
As reported by previous studies cessations are rare phenomenons and double-cessations are much rarer. Brown and Ahlers  \cite{Brown_Ahlers_JFM} observed that cessations account for about $5\%$ of the total reorientations. In our simulations we observe that for $Ra=2\times10^6$ nearly $4$-$7\%$ reorientations are cessation-led. Likewise, for $Ra=2\times10^7$ the percentage of cessations is about $4\%$. However, for the non-rotation case at $Ra=2\times10^7$ cessations account for about $13\%$ for the total reorientations.  
Since our criterion for identification of cessations depends on the amplitude fraction of the Fourier modes ($A_{2}/A_{1}$), and that of reorientations depends on the angular change $|\Delta \Phi_1|$ and angular rotation rate $|\Delta \Phi_1/\Delta t|$, which are independent of each other, some events appear in both the categories while some are exclusive to either cessations or reorientations. Note that the number of reorientations observed in the present study are much lesser compared to the experimental counterparts \cite{Brown_Ahlers_JFM,Brown_Nikolaenko_Ahlers_PRL_2005,Xi_Zhou_Xia_2006} due to the limitations in the length of time histories.

\par
\begin{figure}[h!]
\centering
\includegraphics[scale=0.79]{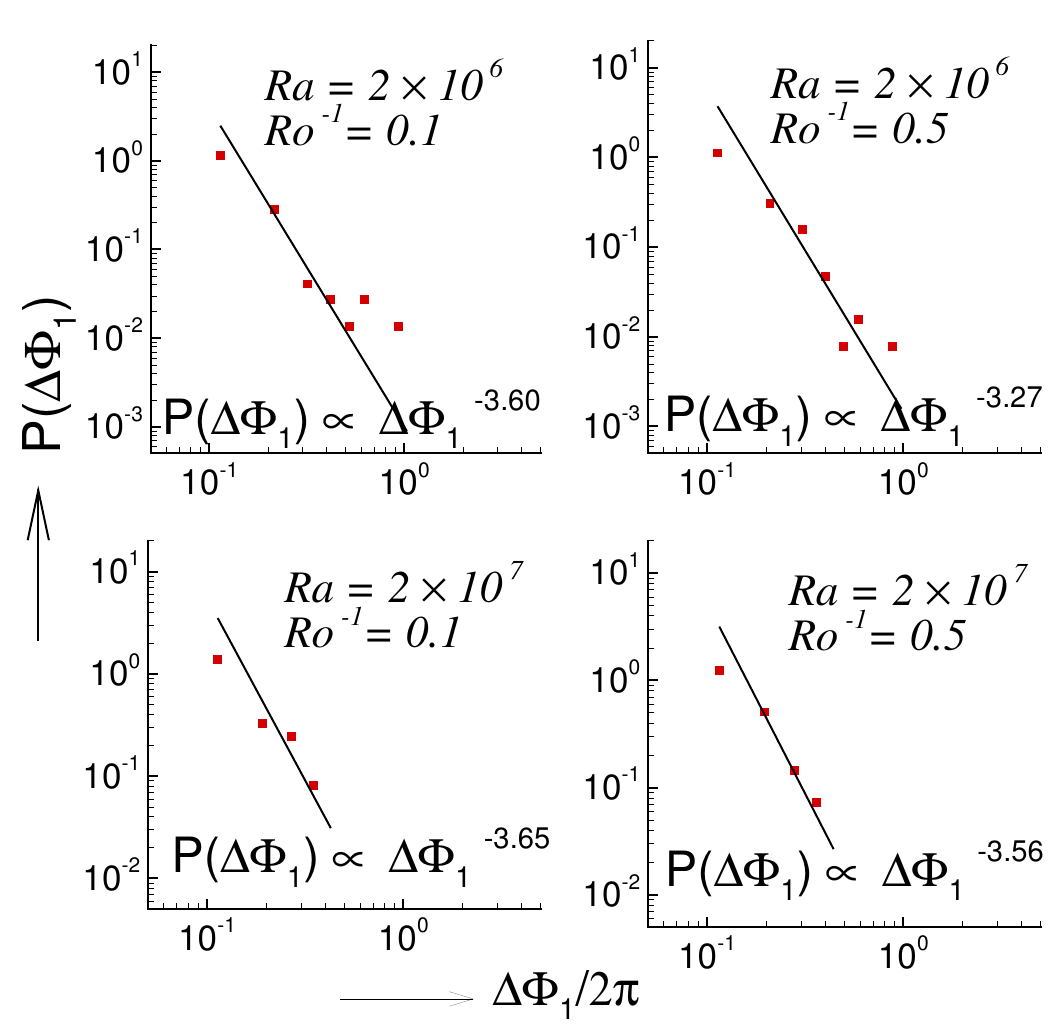}
\caption{PDF of the angular change $\Delta \Phi_1$ for $Ra=2\times10^6$ and $Ra=2\times10^7$  at $Ro^{-1}=0.1$ and $0.5$. Solid line indicates the power law fit by MLE. }
\label{PDF_Reore1}
\end{figure}

Next, we quantify the possibility of occurrence of the reorientations based on the angular change $|\Delta \Phi_1|$.
Figure \ref{PDF_Reore1} shows the probability distribution $P(|\Delta \Phi_1|)$ for different rotation rates.
We observe that the statistics of reorientation follow a power law distribution as $P(|\Delta \Phi_1|) \propto |\Delta \Phi_1|^{-m} $, where $m =3.7\pm 0.5$. The power law fit is carried out by maximum likelihood estimation (MLE) \cite{Powerlaw_Newman_2005,Bevington_book}, to prevent the errors arising due to the binning of data. MLE is preferred over least squares method which can produce substantially inaccurate estimates of parameters for the distribution due to smaller size of the data sample \cite{Bevington_book}. 
In MLE the parameter values of the distribution are calculated such that they maximize the likelihood function, which is the product of probability densities of the individual events. 
The exponent of the power law distribution is calculated as 
\begin{eqnarray}
 m=1+n \left[ \sum_{i=1}^{n}\mathrm{ln}\frac{|\Delta\Phi_{1}^{i}|}{|\Delta\Phi_{1}^{min}|} \right]^{-1}
\end{eqnarray}
where $|\Delta\Phi_{1}^{i}|$, $i=1, ...n$ are the angular changes from $n$ number of observations  and $|\Delta\Phi_{1}^{min}|$ is the minimum angular change. Similar power law fits with negative exponents are observed for all cases (rotating and non-rotating RBC), indicating a monotonically decreasing distribution of $|\Delta \Phi_1|$. This suggests that smaller reorientations are more frequent events, while the larger ones are rare occurrences.
Note that the exponent is well within the range of $-3.25$ to $-4.45$ experimentally observed for different $Ra$ by Brown and Ahlers \cite{Brown_Ahlers_JFM} and  Brown \emph {et al.} \cite{Brown_Nikolaenko_Ahlers_PRL_2005}. The works of Xi and Xia \cite{Xi_Xia_PRE_2008} also found the slope of the power law fit to be of the order of $-3.3$ for unit aspect ratio container. However, all these experiments were on non-rotating RBC. In the present study, we infer that even for rotating RBC, the reorientations follow a power law distribution.

 \begin{figure}[h!]
\centering
\includegraphics[scale=0.80]{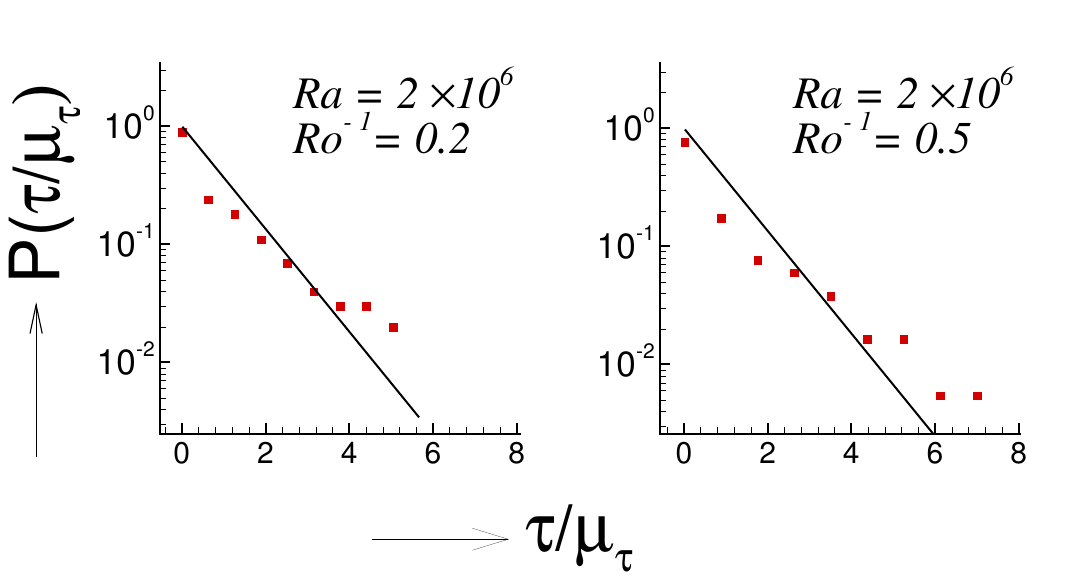}
\caption{PDF of time interval between successive reorientations for $Ra=2\times10^6$ at $Ro^{-1}=0.2$ and $0.5$. The solid line indicates the exponential function $P(\uptau/\mu_{\uptau})$~=$~\exp(\uptau/\mu_{\uptau})$. }
\label{PDF_Tau}
\end{figure}
\par
We now evaluate the occurrence of reorientations with time. For this we consider the time interval between two successive reorientations ($\tau$). Experimental studies by  Xi and Xia \cite{Xi_Xia_PRE_2008}, Brown and Ahlers \cite{Brown_Ahlers_JFM} and  Brown \emph {et al.} \cite{Brown_Nikolaenko_Ahlers_PRL_2005} have reported that the PDF of time interval of reorientations can be well fitted with an exponential function, which suggests that the occurrence of reorientations follows a Poisson process. 
Figure \ref{PDF_Tau} shows the probability distribution $P(\uptau/\mu_{\uptau})$ for different rotation rates ($Ro^{-1}=0.2$ and $0.5$), where the solid line shows the exponential function $P(\uptau/\mu_{\uptau})~$=$~\exp(\uptau/\mu_{\uptau})$ with $\mu_{\uptau}$ the average time interval of reorientations. 
Note that for $Ra=2\times10^6$ the time interval is in good agreement with the exponential function, while deviations are observed in $Ra=2\times10^7$, which might be due to the lesser number of observations. Poisson distribution of the time interval suggests that reorientations are independent of each other and they do not have any memory of the previous event \cite{Xi_Xia_PRE_2008}.

\section{Conclusions}\label{Sec:Conclusion}
We have presented a detailed three-dimensional numerical investigation on the dynamics and statistics of reorientations of LSC in turbulent RRBC in a unit aspect ratio cylindrical cell  with $Pr=0.7$ and two different Rayleigh numbers $Ra=2\times10^6$ and $2\times10^7$, for a wide range of rotation rate ($0\leq Ro^{-1}\leq 30$). We have been able to identify different flow regimes based on the energy contained in the Fourier modes computed for the field recorded in the azimuthal direction of the mid-plane of the container. We find LSC dominated flow at the low rotation rate ( $Ro^{-1}\lesssim1$) while presence of multiple roll structures like quadrupole and sextupole have been observed at higher rotation rates. Presence of LSC significantly affects the overall heat transfer and the boundary layer thickness. Along the direction of LSC, the thermal boundary layer width $\delta_{\theta}$ shows an asymmetric trend, as it is thicker at one side and thinner near the opposite side of the lateral wall. 
However, perpendicular to LSC, the boundary layer thickness varies almost symmetrically.
\par
Further, the reorientations of LSC have been characterized as rotation-led and cessation-led based on their nature of occurrence, and as partial and complete reversal depending on the degree of phase change. In addition to the previously reported rare events like cessations and double-cessation, an interesting event of multiple-cessation have been observed, where more than two cessations occur in quick succession. Cessation-led reorientations are rare events, which account for about $4$-$7\%$ of the total number of reorientations. Double-cessations are even rarer, and multiple-cessations are the rarest of all reorientations. 
At moderate rotation rates, significant increase in the number of rotation-led reorientations are observed, while cessation-led ones are unaffected by rotation. Similar to non-rotating RBC, we find that the probability distribution of the reorientations exhibits a power law distribution with the exponent nearly equal to $-3.7$. In addition the PDF of time interval between two successive reorientations follows a Poisson distribution, which indicates that reorientations are independent of each other.

\section*{Acknowledgement}
All simulations have been carried out in the `Param-Ishan' computing facility at IIT Guwahati.
 \section*{References}
\bibliography{reference}

\end{document}